\documentclass[aps,prl,twocolumn,superscriptaddress]{revtex4}
\pdfoutput=1
\usepackage[latin9]{inputenc}
\usepackage{graphicx,epsf}
\usepackage{amssymb}
\usepackage{amsmath}
\usepackage{color}
\usepackage{bm}

\newcommand{\qph}{\quad \phantom{.}}
\newcommand{\qqph}{\qquad \phantom{.}}
\newcommand{\w}{\omega}
\newcommand{\onebm}{SBM1}
\newcommand{\twobm}{SBM2}
\newcommand{\spinboson}{{\rm sb}}
\newcommand{\critical}{{\rm c}}
\newcommand{\bath}{{\rm bath}}
\newcommand{\optimal}{{\rm opt}}
\newcommand{\spin}{\sigma}
\newcommand{\eff}{{\rm eff}}

\RequirePackage[
   hyperindex,colorlinks,bookmarksnumbered,
   plainpages=true, pdfstartview=FitH,bookmarks=false
]{hyperref}
\hypersetup{linkcolor=blue,urlcolor=blue,citecolor=blue}

\definecolor{darkgreen}{rgb}{0,0.5,0}
\definecolor{darkblue}{rgb}{0,0,0.5}
\definecolor{purple}{rgb}{0.35,0,0.35}
\definecolor{orange}{rgb}{1,0.5,0}
\definecolor{wbcolor}{rgb}{0,.6,1}
\definecolor{todocolor}{rgb}{1,0,0}
\definecolor{jvdcolor}{rgb}{0,0,1}

\newcommand{\Eq}[1]{Eq.~(\ref{#1})}
\newcommand{\Eqs}[1]{Eqs.~(\ref{#1})}

\newcommand{\Ref}[1]{Ref.~\onlinecite{#1}}
\newcommand{\Refs}[1]{Refs.~\onlinecite{#1}}
\newcommand{\Fig}[1]{Fig.~\ref{#1}}

\newcommand{\pdag}{{\phantom{\dagger}}}

\begin{document}
%%%%%%%%%%%%%%%%%%%%%%%%%%%%%%%%%%%%%%%%%%%%%%%%%%%%%%%%%%%%%%%%%%%%%%%
\title{
Critical and strong-coupling phases in one- and two-bath spin-boson models}

\author{Cheng Guo}
\author{Andreas Weichselbaum}
\author{Jan von Delft}
\affiliation{Ludwig-Maximilians-Universit\"at M\"unchen, Germany}
\author{Matthias Vojta}
\affiliation{Institut f\"ur Theoretische Physik, Technische Universit\"at Dresden, 01062 Dresden, Germany}

\date{\today}

\begin{abstract}
  For phase transitions in dissipative quantum impurity models, the
  existence of a quantum-to-classical correspondence has been
  discussed extensively.  We introduce a variational matrix product
  state approach involving an optimized boson basis, rendering
  possible high-accuracy numerical studies across the entire phase
  diagram.
  For the sub-ohmic spin-boson model with a power-law bath spectrum
  $\propto \w^s$, we confirm classical mean-field behavior for
  $s<1/2$, correcting earlier numerical renormalization-group results.
  We also provide the first results for an XY-symmetric
  model of a spin coupled to two competing bosonic baths, where we
  find a rich phase diagram, including both critical and
  strong-coupling phases for $s<1$, different from that of classical
  spin chains.
This illustrates that symmetries are decisive for whether or not
  a quantum-to-classical correspondence exists.
\end{abstract}


\maketitle

%%%%%%%%%%%%%%%%%%%%%%%%%%%%%%%%%%%%%%%%%%%%%%%%%%%%%%%%%%%%%%%%%%%%%%%

Quantum spins in a bosonic environment are model systems in diverse
areas of physics, ranging from dissipative quantum mechanics to
impurities in magnets and biological systems \cite{leggett}.
In this paper we
consider the spin-boson model and a generalization
thereof to two baths, described by $\mathcal{H}_\spinboson =
- \vec{h} \cdot \vec\sigma/ 2 + \mathcal{H}_\bath$, with
\begin{eqnarray}
\label{eq:h}
\mathcal{H}_\bath
&=&
   \sum_{i = x,y} \sum_{q} \left[
\w_q \hat{B}_{q i}^{\dagger} \hat{B}_{q i}^\pdag +
\lambda_{q i} \frac{\sigma_{i}}{2}
( \hat{B}^\pdag_{q i} + \hat{B}_{q i}^{\dagger} )\right] .
\qph
\end{eqnarray}
The two-level system (or quantum spin, with $\sigma_{x,y,z}$ being the
vector of Pauli matrices) is coupled both to an external field
$\vec{h}$ and, via $\sigma_x$ and $\sigma_y$, to two independent
bosonic baths, whose spectral densities $J_i (\omega) = \pi \sum_{q}
\lambda_{q i}^{2} \delta(\w -\w_q)$ are assumed to be of power-law
form:
\begin{equation}
J_{i}(\w) = 2\pi\, \alpha_{i}\, \omega_{\rm c}^{1-s} \, \omega^s\,, \quad
 0<\w<\w_{\rm c} =1 \; .
\label{power}
\end{equation}
Such models are governed by the competition between the local field, which tends to point
the spin in the $\vec{h}$ direction, and the dissipative effects of the bosonic
baths.

Indeed, the standard one-bath spin-boson model (\onebm), obtained for
$\alpha_y=h_y=0$, exhibits an interesting and much-studied
\cite{leggett, kehrein_spin-boson_1996,bulla_numerical_2003,VTB,fehske,winter_quantum_2009,wong_density_2008}
quantum phase transition (QPT) from a delocalized to a localized
phase, with $\langle \sigma_x \rangle = 0$ or $\neq 0$, respectively,
as $\alpha_x$ is increased past a critical coupling
$\alpha_{x, \critical}$.
According to statistical-mechanics arguments, this transition is in the same
universality class as the thermal phase transition of the
one-dimensional (1D) Ising model with $1/r^{1+s}$ interactions.  This
quantum-to-classical correspondence (QCC) predicts mean-field
exponents for $s<1/2$, where the Ising model is above its
upper-critical dimension
\cite{luijten_classical_1997,fisher_critical_1972}.

Checking this prediction numerically turned out to be challenging.
Numerical renormalization-group (NRG) studies of \onebm\ yielded
non-mean-field exponents for $s<1/2$ \cite{VTB}, thereby seemingly
negating the validity of the QCC. However, the authors of
Ref.~\onlinecite{VTB} subsequently concluded
\cite{vojta_erratum:_2009} that those results were not reliable, due to two inherent
limitations of NRG, which they termed (i) Hilbert-space truncation and
(ii) mass flow.
Problem (i) causes errors for critical exponents that characterize the
flow into the localized phase at zero temperature, since $\langle
\sigma_x\rangle \neq 0$ induces shifts in the bosonic displacements
$\hat{X}_{q} = (\hat{B}^\pdag_{q} + \hat{B}_{q}^\dagger)/\sqrt{2}$
of the bath oscillators which diverge in the low-energy limit for $s<1$ and hence
cannot be adequately described in the truncated boson Hilbert space
used by NRG \cite{bulla_numerical_2005}.
Problem (ii) arises for non-zero temperatures, due to NRG's neglect of
low-lying bath modes with energy smaller than temperature
\cite{flow}. In contrast to NRG, two  recent numerical studies of
\onebm, using Monte Carlo methods \cite{winter_quantum_2009} or a
sparse polynomial basis \cite{fehske}, found mean-field exponents in
agreement with the QCC.  Nevertheless, other recent works
continue to advocate the failure of the QCC \cite{si09}.

The purpose of this Letter is twofold.
First, we show how the problem (i) of Hilbert-space truncation can be
controlled systematically by using a variational matrix product state
(VMPS) approach formulated on a Wilson chain.  The key idea is to
variationally construct an optimized boson basis (OBB) that captures
the bosonic shifts induced by $\langle \sigma_x \rangle \neq 0$.  The
VMPS results confirm the predictions of the QCC for the QPT of \onebm\
at $T=0$.  (Problem (ii) is beyond the scope of this work.)
Second, we use the VMPS approach to study an XY-symmetric version of the two-bath
spin-boson model (\twobm), with $\alpha_x = \alpha_y$. This model
arises, e.g., in the contexts of impurities in quantum magnets \cite{antonio_xy,rg_bfk}
and of noisy qubits \cite{antonio_xy,dima}, and displays the phenomenon of ``frustration
of decoherence'' \cite{antonio_xy}: the two baths compete (rather than cooperate), each
tending to localize a different component of the spin.  As a result, a non-trivial
intermediate-coupling (i.e.  critical) phase has been proposed to emerge for $s<1$
\cite{rg_bfk}, which has no classical analogue. To date, the existence of this phase
could only be established in an expansion in $(1-s)$, and no numerical results are
available. Here we numerically investigate the phase diagram, and,
  surprisingly,  find that the perturbative predictions are valid
  for a small range of $s$ and $\alpha$ only. We conclusively
  demonstrate the absence of a QCC for this model.

%%%%%%%%%%%%%%%%%%%%%%%%%%%%%%%%%%%%%%%%%%%%%%%%%%%%%%%%%%%%%%%%%%%%%%%

\emph{Wilson chain.---} Following
Refs.~\onlinecite{bulla_numerical_2003,bulla_numerical_2005}, which
adapted Wilson's NRG to a bosonic bath, we discretize the latter using
a logarithmic grid of frequencies $\omega_{k i} \propto \Lambda^{-k}$
(with $\Lambda > 1$ and $k$ a positive integer) and map
$\mathcal{H}_\bath$ onto a so-called Wilson chain of $(L-1)$ bosonic sites:
\begin{eqnarray}
\label{eq:Hsb_chain}
\mathcal{H}^{(L-1)}_\bath & = & \sum_{i=x,y} \Biggl[
\sqrt{\frac{\eta_{i}}{\pi}}\frac{\sigma_{i}}{2}
(\hat{b}^\pdag_{1 i} + \hat{b}_{1 i}^{\dagger}) \Biggr.
\\ & & \Biggl. + \sum_{k=1}^{L-2} t_{k i}(\hat{b}_{k i}^{\dagger}
\hat{b}^\pdag_{k+1,i}+h.c.)
+ \epsilon_{k i} \hat{n}_{k i}  \Biggr] \; .
\qqph
\nonumber
\end{eqnarray}
Here $\hat{n}_{k i} = \hat{b}_{k i}^{\dagger}\hat{b}^\pdag_{k i}$,
with eigenvalue $n_{k i}$, counts the bosons of type $i$ on chain site
$k$; the detailed form of the hopping parameters $t_{k i}$, on-site
energies $\epsilon_{k i}$ (both $\propto \Lambda^{-k}$), and the
coupling $\eta_i$ between spin component $\sigma_{i}$ and
site 1, are obtained following \Refs{zitko_energy_2009,SupplInfo}.
To render a numerical treatment feasible, the infinite-dimensional
bosonic Hilbert space at each site $k$ is truncated by restricting the
boson number to $0 \le n_{k i} < d_k$ ($d_k \le 14$ in
Refs.~\onlinecite{bulla_numerical_2003,bulla_numerical_2005}).

\begin{figure}
\includegraphics[width=8.5cm]{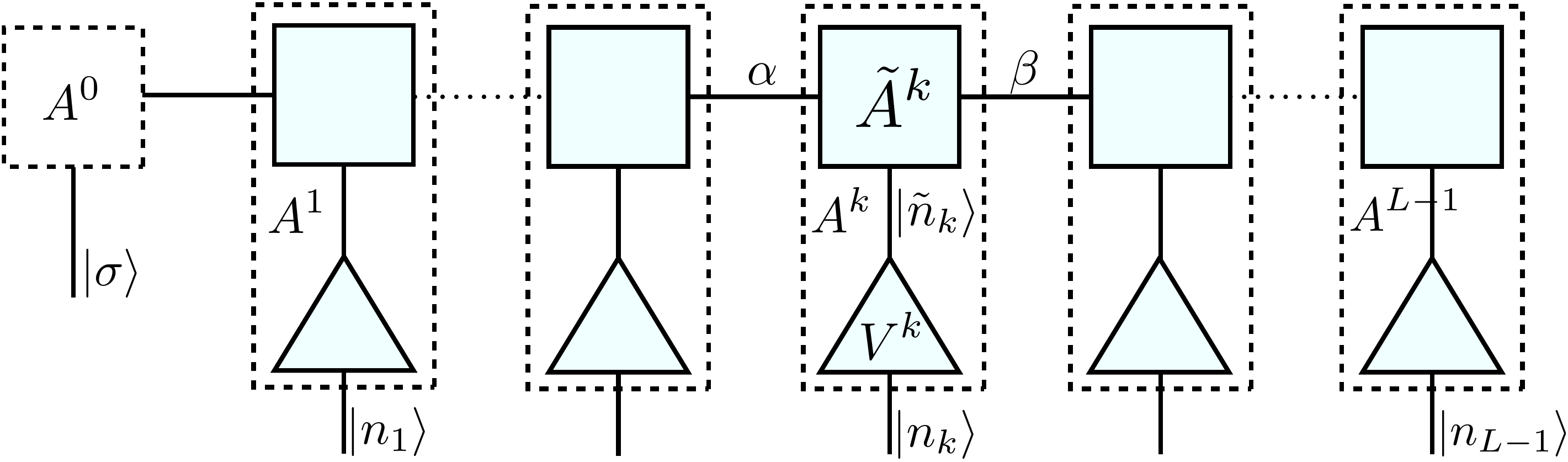}
\caption{
Depiction of the MPS
\Eq{eq:mps}, with each $A$-matrix expressed
in an optimal boson basis via $A=  \tilde A V$
[\Eq{eq:A_decomposition}].
}
\label{fig:mps}
\end{figure}

The standard NRG strategy for finding the ground state of
$\mathcal{H}_\spinboson^{(L)} = - \vec h \cdot \vec \sigma/2 +
\mathcal{H}^{(L-1)}_\bath$ is to iteratively diagonalize it one site at a
time, keeping only the lowest-lying $D$ energy eigenstates at each
iteration.  This yields a $L$-site matrix-product state (MPS)
\cite{weichselbaum_variational_2009,saberi_matrix-product-state_2008,%
schollwock_2011}
of the following form (depicted in \Fig{fig:mps}, dashed boxes):
\begin{equation}
  |G\rangle= \sum_{\spin = \uparrow, \downarrow} \sum_{ \{ \vec n \}}
  A^0[\spin] A^{1}[n_{1}] \cdots A^{L-1}[n_{L-1}] |\spin \rangle |\vec n
  \rangle \; .
\label{eq:mps}
\end{equation}
Here $|\spin\rangle = |\! \uparrow\rangle$, $|\! \downarrow\rangle$
are eigenstates of $\sigma_{x}$; the states $|\vec n\rangle = |n_1,
\dots, n_{L-1} \rangle$ form a basis of boson-number eigenstates
within the truncated Fock space, with $\hat{n}_{k i} |\vec n\rangle =
n_{k i} |\vec n\rangle$ and $0 \le n_{k i} < d_k$.  For \twobm,
$n_k = (n_{k x}, n_{k y})$ labels the states of supersite $k$
representing both chains.  Each $A^k[n_k]$ is a matrix (not
necessarily square, but of maximal dimension $D\times D$, with $A^0$ a
row matrix and $A^{L-1}$ a column matrix), with matrix elements
$\left(A^k[n_k]\right)_{\alpha \beta}$.

The need for Hilbert-space truncation with small $d_k$ prevents NRG
from accurately representing the shifts in the displacements
$\hat{x}_{k i} = (\hat{b}^\pdag_{k i} + \hat{b}^\dagger_{k i})/\sqrt{2} $ that occur in the localized phase. This
problem can be avoided, in principle, by using an OBB,
chosen such that it optimally represents the quantum fluctuations of
\emph{shifted} oscillators, $\hat{x}'_{k i} = \hat{x}_{k i} -
\langle \hat{x}_{k i} \rangle$.  While attempts to
accommodate this strategy within standard NRG were unsuccessful
\cite{bulla_numerical_2005}, it was shown to work well \cite{fehske}
using an alternative representation of \onebm\ using a sparse
polynomial basis.

%%%%%%%%%%%%%%%%%%%%%%%%%%%%%%%%%%%%%%%%%%%%%%%%%%%%%%%%%%%%%%%%%%%%%%%

\emph{VMPS method.---} We now show that an OBB can
also be constructed on a Wilson chain. To this end, view the state
$|G\rangle$ of \Eq{eq:mps} as a MPS ansatz for the ground state of
$\mathcal{H}_\spinboson^{(L)}$, that is to be optimized \emph{variationally}
using standard MPS methods
\cite{weichselbaum_variational_2009,saberi_matrix-product-state_2008,schollwock_2011}. To
allow the possibility of large bosonic shifts, we represent the
$A$-matrix elements as
\cite{zhang_density_1998,weie_optimized_2000,%
nishiyama_numerical_1999} (\Fig{fig:mps}, solid lines)
\begin{equation}
(A^k[n_k])_{\alpha \beta} = \sum_{\tilde
  n_k = 0}^{d_\optimal-1} (\tilde A^k[\tilde n_k])_{\alpha \beta}
V^k_{\tilde n_k n_k} \quad (k \ge 1) \; .
\label{eq:A_decomposition}
\end{equation}
Here $V^k$ in effect implements a transformation to a new boson basis
on site $k$, the OBB, of the form $|\tilde n_k \rangle = \sum_{n_k =
  0}^{d_k-1} V^k_{\tilde n_k n_k} |n_k \rangle$ with $0 \le \tilde n_k <
d_\optimal$.  (For SBM2, $V^k$ is a rank-3 tensor.)
This ansatz has the advantage that the
size of the OBB,  $d_\optimal$, can be chosen to be much smaller ($d_{\rm opt}
  \lesssim 50$) than $d_k$.
Following standard DMRG strategy, we optimize the $\tilde A^k$ and
$V^k$ matrices one site at a time through a series of variational
sweeps through the Wilson chain. As further possible improvement
before optimizing a given site,
the requisite boson shift can be implemented by hand in the Hamiltonian
itself: we first determine
the ``current'' value of the bosonic shift $\langle \hat{x}_{k i}
\rangle$ using the current variational state $|G\rangle$,
then use it as starting point to variationally optimize
a new $|G'\rangle$
with respect to the shifted Hamiltonian ${\cal H}_\spinboson^{\prime (L)}
(\hat{b}_{k i},
\hat{b}_{k i}^{\dagger}) = {\cal H}_\spinboson^{(L)}(\hat{b}_{k i}', \hat{b}_{k i}^{\prime
  \dagger})$, with $\hat{b}'_{k i} = \hat{b}_{k i} - \langle
\hat{x}_{k i} \rangle/\sqrt 2$.
The shifted OBB protocol, described in detail in \Ref{SupplInfo},
allows shifts that would have required
$d_k^\eff \approx 10^{10}$ states in the original boson basis
to be treated using rather small $d_k$ (we used $d_k = 100$).

%%%%%%%%%%%%%%%%%%%%%%%%%%%%%%%%%%%%%%%%%%%%%%%%%%%%%%%%%%%%%%%%%%%%%%%

\emph{Spin-boson model. ---}
We applied the VMPS method to \onebm\ ($\alpha_y =h_y = 0$), with
dissipation strength $\alpha \equiv \alpha_x$ and fixed transverse
field $h_z=0.1$, at $T=0$.  We focussed on the QPT between the delocalized and
localized phases in the subohmic case, $s<1$. Here, the controversy
\cite{VTB,vojta_erratum:_2009,winter_quantum_2009,fehske,si09}
concerns the order-parameter exponents $\beta$ and $\delta$, defined
via $\langle \sigma_x \rangle \propto (\alpha-\alpha_\critical)^\beta$
at $h_x=0$ and $\langle \sigma_x \rangle \propto h_x^{1/\delta}$ at
$\alpha =\alpha_\critical$, respectively. QCC predicts mean-field
values $\beta_{\rm MF}=1/2$, $\delta_{\rm MF}=3$ for $s<1/2$
\cite{fisher_critical_1972}, whereas initial NRG results \cite{VTB}
showed $s$-dependent non-mean-field exponents.

\begin{figure}
\center
\includegraphics[width=8.7cm,clip]{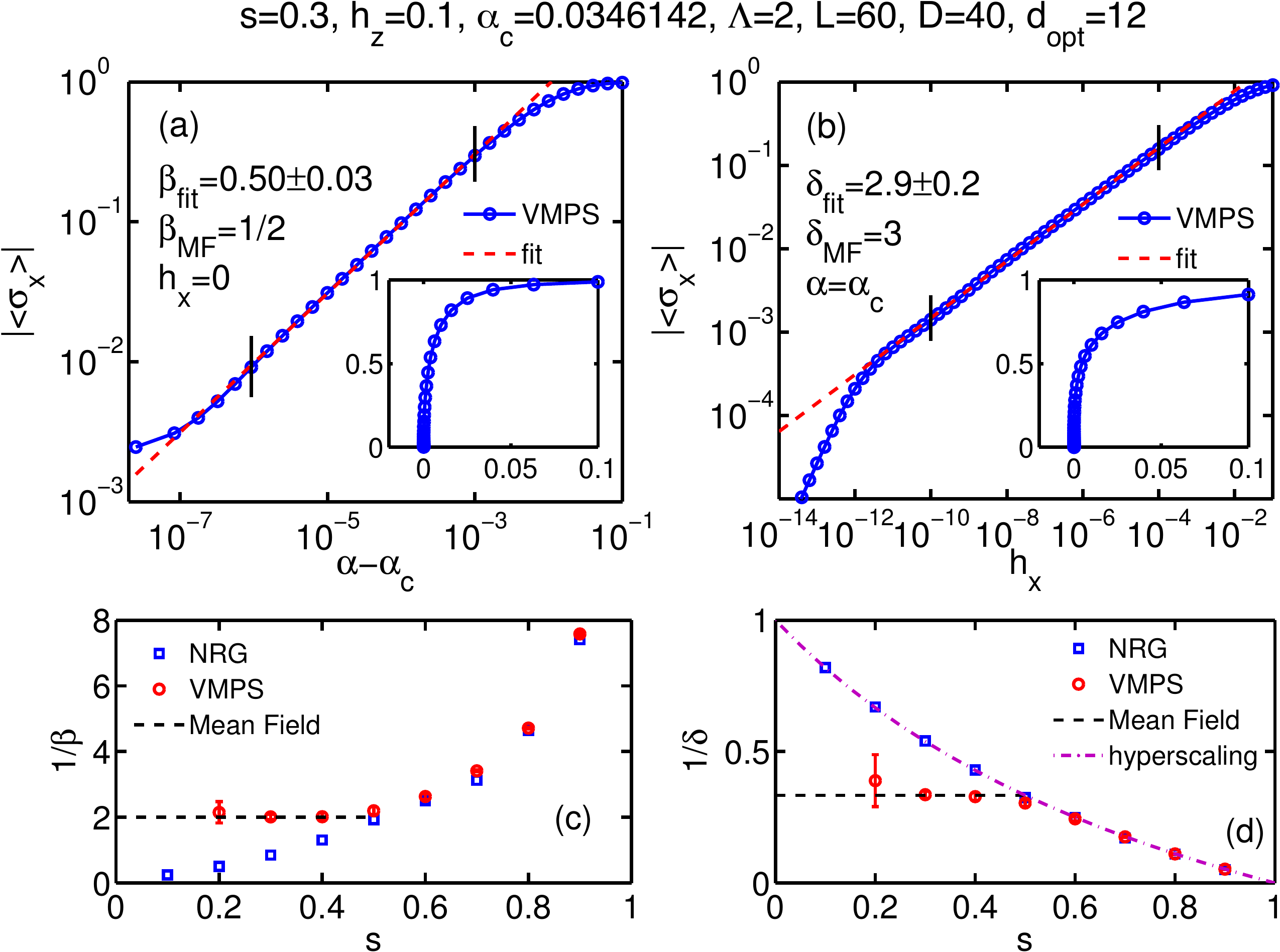}
\caption{
VMPS results for the order parameter of \onebm\ near criticality.
a) $\langle \sigma_x \rangle$ vs.\ $(\alpha-\alpha_\critical)$
at $h_x = 0$, and b) $\langle \sigma_x \rangle$ vs.\ $h_x$ at
$\alpha =\alpha_\critical$,
on linear plots (insets) or log-log plots (main panels).
Dashed lines are  power-law fits in the ranges between
the vertical marks.
c,d) Comparison of the exponents $\beta$ and $\delta$
for different $s$ obtained from VMPS, NRG \cite{VTB}, mean-field theory,
and, in  d), the exact hyperscaling result $\delta=(1+s)/(1-s)$
which applies for $s>1/2$. (See also \cite{SupplInfo}, Fig.~S7).
}
\label{fig:fit}
\end{figure}

In Fig.~\ref{fig:fit}a, we show sample VMPS results for $\langle
\sigma_x \rangle$ vs.  $(\alpha -\alpha_\critical)$ for $s=0.3$ at
$h_x = 0$, where $\alpha_\critical$ was tuned to yield the best
straight line on a log-log plot.  The results display power-law
behavior over more than 3 decades, with an exponent
$\beta = 0.50 \pm 0.03$.
Deviations at small $(\alpha -\alpha_\critical)$ can be
attributed to a combination of finite chain length and numerical
errors of VMPS. Fig.~\ref{fig:fit}b shows $\langle \sigma_x \rangle$
vs. $h_x$ at $\alpha = \alpha_\critical$, and a power-law fit over 6 decades
results in $\delta=2.9 \pm 0.2$.
Power laws of similar quality can be obtained for all $s\gtrsim 0.2$
\cite{SupplInfo,smalls_foot} (see \cite{SupplInfo}, Fig.~S7).

The exponents $\beta$ and $\delta$ obtained from such fits are
summarized in Figs.~\ref{fig:fit}c,d. For $s<1/2$ they are consistent with
the mean-field values predicted by QCC, also found in Monte-Carlo
\cite{winter_quantum_2009} and exact-diagonalization studies
\cite{fehske}, but are at variance with the NRG data of
Ref.~\onlinecite{VTB}.  Since both NRG and VMPS handle the same
microscopic model $\mathcal{H}_\spinboson^{(L)}$ defined on the Wilson
chain, but VMPS can deal with much larger $d^\eff_k$ values
($\lesssim 10^{10}$ in \Fig{fig:fit}) than NRG, the incorrect NRG
results must originate from Hilbert-space truncation, as anticipated
in Ref.~\onlinecite{vojta_erratum:_2009}.
Indeed, artificially restricting $d_k$ to small values in VMPS reproduces the
incorrect NRG exponents (see \cite{SupplInfo}, Fig.~S6).

%%%%%%%%%%%%%%%%%%%%%%%%%%%%%%%%%%%%%%%%%%%%%%%%%%%%%%%%%%%%%%%%%%%%%%%

\emph{Two-bath model. ---}
We now turn to \twobm, a generalization of the spin-boson
model. Here, the two baths may represent distinct noise sources
\cite{antonio_xy,dima} or XY-symmetric magnetic fluctuations
\cite{antonio_xy,rg_bfk,si_transistor}. Perturbation theory shows that
the two baths compete: A straightforward expansion around the
free-spin fixed point ($\alpha=h=0$) results in the following one-loop
renormalization-group (RG) equations at $\vec h = 0$:
\begin{equation}
\beta(\alpha_x) = (1-s) \alpha_x - \alpha_x \alpha_y,~~
\beta(\alpha_y) = (1-s) \alpha_y - \alpha_x \alpha_y .
\label{eq:rg}
\end{equation}
For $\alpha \! \equiv \! \alpha_x \! = \! \alpha_y$, these equations
predict a stable inter\-mediate-coupling fixed point at $\alpha^\ast =
1-s$, describing a {\em critical} phase. It is characterized by
$\langle \vec{\sigma} \rangle =0$, a \emph{non-linear} response of
$\langle \vec \sigma \rangle$ to an applied field $\vec h$, and a
finite ground-state entropy smaller than $\ln 2$, all corresponding to
a fluctuating fractional spin \cite{rg_bfk,VBS}. This phase is
unstable w.r.t.\ finite bath asymmetry ($\alpha_x\neq\alpha_y$) and
finite field. It had been  assumed \cite{rg_bfk} that this
critical phase exists for all $0<s<1$ and is reached for any $\alpha$.

\begin{figure}
\includegraphics[width=8.5cm,clip]{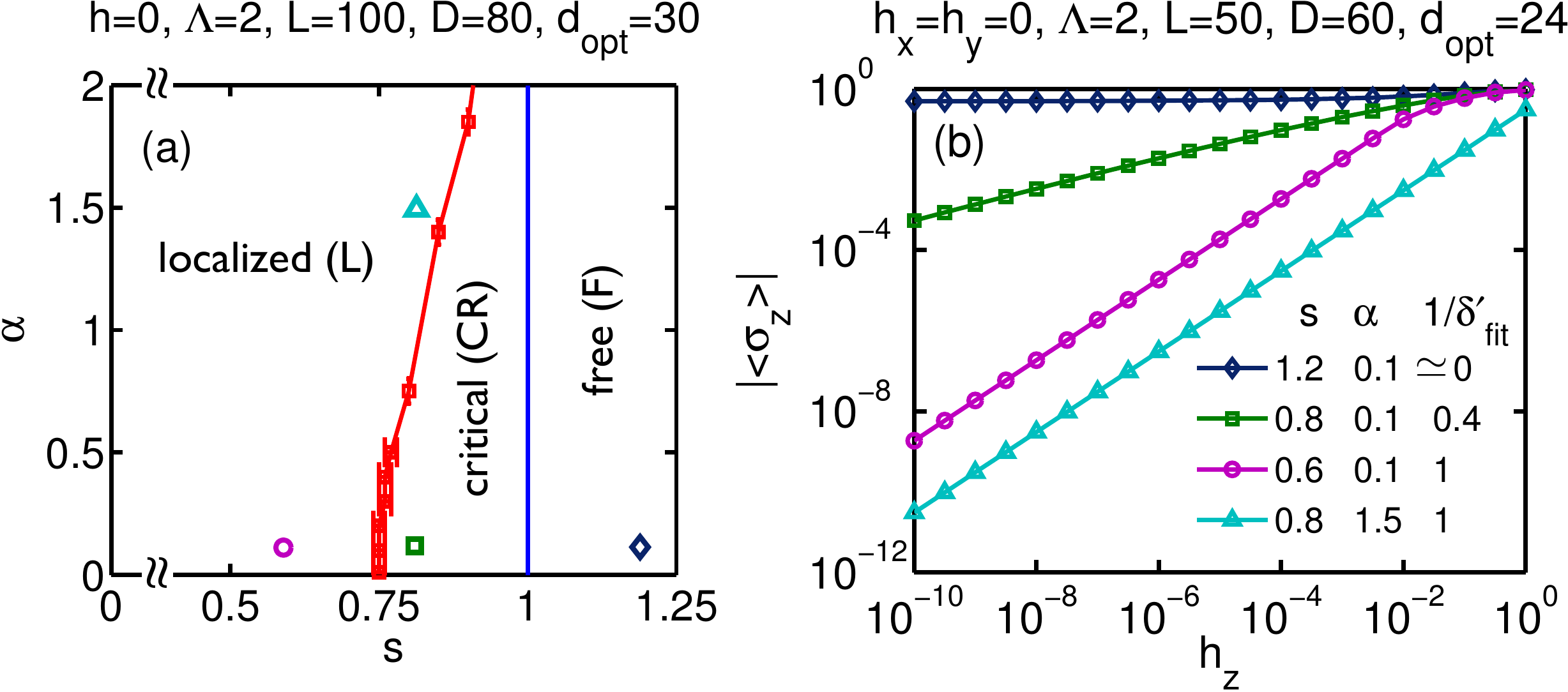}
\caption{
  a) Phase diagram of \twobm\ in the $s$--$\alpha$ plane for
  $\vec{h}=0$, with dissipation strenght
  $\alpha\equiv\alpha_x=\alpha_y$.  The critical phase only exists for
  $s^\ast<s<1$, and its boundary $\alpha_\critical \to \infty$ for $s
  \to 1^-$.
(\Ref{SupplInfo}
describes the determination of the phase boundary and
gives a 3D sketch of the  $s$-$\alpha$-$h_z$ phase diagram,
see Fig.~S8.)
b) Tranverse-field response of \twobm, $\langle \sigma_z \rangle
$ $\propto h_z^{1/\delta'}$, for four choices of $s$ and $\alpha$,
showing free (diamonds), critical (squares) and localized
(triangles,circles) behavior.
}
\label{fig:sbm2res} \vspace{-6mm}
\end{figure}

\begin{figure}[!b]
\center
\includegraphics[width=8cm,clip]{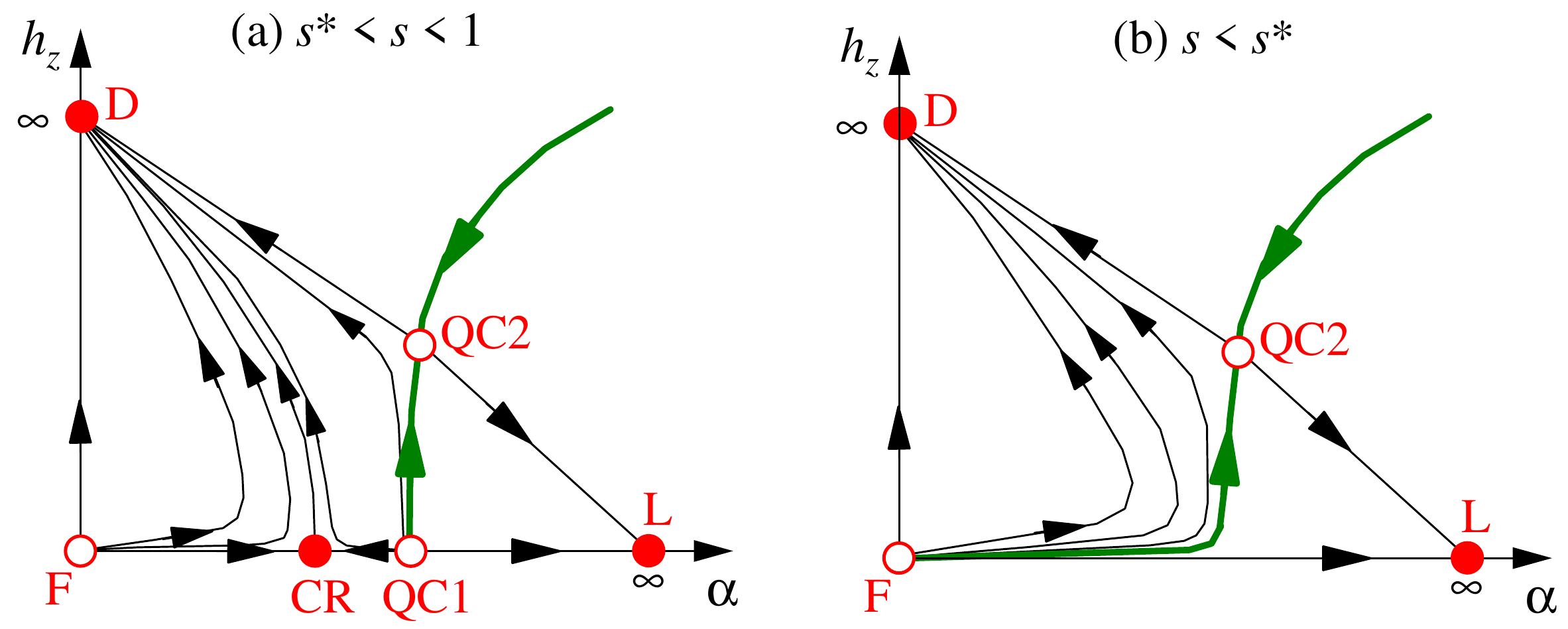}
\caption{
Schematic RG flow for \twobm\ in the $\alpha$-$h_z$ plane ($h_x=h_y=0$).
The thick lines correspond to continuous QPT;
the full (open) circles are stable (unstable) fixed points,
for labels see text.
a) $s^\ast<s<1$: CR is reached for small $\alpha$ and $h_z=0$, it is
separated from L by a QPT controlled by the multicritical QC1 fixed
point. Eq.~\eqref{eq:rg} implies that CR is located at
  $\alpha^\ast = 1-s + \mathcal{O}[(1-s)^2]$. For finite $h_z$, a QPT
between D and L occurs, controlled by QC2.
b) $0<s<s^\ast$: both CR and QC1 have disappeared, such that the only
transition is between D and L.
}
\label{fig:flow}
\end{figure}

We have extensively studied \twobm\ using VMPS; the results are
summarized in the $\vec h=0$ phase diagram in Fig.~\ref{fig:sbm2res}a
and the flow diagrams in Fig.~\ref{fig:flow}.
Most importantly, we find that the critical phase (CR) indeed exists,
but only for $s^\ast < s < 1$, with a universal $s^\ast = 0.75\pm0.01$. Even in
this $s$ range, the critical phase is left once $\alpha$
is increased beyond a critical value $\alpha_{\rm c}(s)$, which marks
the location of a continuous QPT into a localized phase (L)
with spontaneously broken XY symmetry and finite $\langle
\sigma_{x,y}\rangle$. This localized phase exists down to $s=0$,
Fig.~\ref{fig:sbm2res}a.  It can be destabilized by applying a
transverse field $h_z$ beyond a critical value
$h_{z}^\critical(\alpha)$, marking the location of a
continuous QPT into a delocalized phase (D)
with unique ground state (see Ref.~\cite{SupplInfo},
Fig.~S9).  Finally, for $s\geq 1$ we only find
weak-coupling behavior, i.e., the impurity behaves as a free spin (F).

In Fig.~\ref{fig:sbm2res}b (and Ref.~\cite{SupplInfo}, Fig.~S10)
we show results for the transverse-field response, $\langle \sigma_z
\rangle \propto h_z^{1/\delta'}$, which can be used to characterize
the different zero-field phases.
 $\langle \sigma_z \rangle$ is linear
in $h_z$ in L $(\delta' = 1)$, sublinear in CR $(\delta' > 1)$, and
extrapolates to a finite value in F. For CR, a perturbative
calculation gives $1/\delta'=(1-s) + \mathcal{O}([1-s]^2)$
\cite{rg_bfk} (confirmed numerically in Ref.~\cite{SupplInfo},
  Fig.~S11b), while the linear response in L corresponds to that of
an ordered XY magnet to a field perpendicular to the easy plane.

From the VMPS results, we can schematically construct the RG flow,
Fig.~\ref{fig:flow}.  There are three stable RG fixed points for
$s^\ast < s < 1$, corresponding to the L, D, and CR phases. From this
we deduce the existence of two unstable critical fixed points, QC1 and
QC2, controlling the QPTs (Fig.~\ref{fig:flow}a). Eq.~\eqref{eq:rg}
predicts that, as $s\to 1^-$, CR merges with F; this is consistent
with our results for $\delta'$ which indicate $\delta'\to\infty$
  as $s\to 1^-$ (Ref.~\onlinecite{SupplInfo}, Fig.~S11b).
The behavior of the phase boundary $\alpha_\critical$ in
Fig.~\ref{fig:sbm2res}a suggests that QC1 moves towards
$\alpha=\infty$ for $s\to 1^-$. Thus, for $s\geq 1$ only F is stable
on the $\vec{h}=0$ axis.
Conversely, from Eq.~\eqref{eq:rg} and Fig.~\ref{fig:sbm2res}a we
  extract that, upon lowering $s$, CR (QC1) moves to larger (smaller)
  $\alpha$.
  From the absence of CR for small $s$ we then conclude that CR and
  QC1 merge and disappear as $s\to {s^\ast}^+$. Consequently, for
  $s<s^\ast$ we have only D and L as stable phases, separated by a
  transition controlled by QC2, Fig.~\ref{fig:flow}b. The merger of CR
  and QC1 at $s=s^\ast$ also implies that the phase boundary between
  CR and L in Fig.~\ref{fig:sbm2res}b at $s^\ast$ is vertical at small
  $\alpha$ (Ref.~\cite{SupplInfo}, Sec.~V.C), because the
  merging point on the $\alpha$ axis defines the finite value of
  $\alpha_{\rm c}$ at $s\to {s^\ast}^+$.

    Taken together, the physics of \twobm\ is much richer than that of
    a classical XY-symmetric spin chain with long-range interactions,
    which only shows a single thermal phase transition
    \cite{koster}. Given this apparent failure of QCC for \twobm, it
    is useful to recall the arguments for QCC for \onebm: A Feynman
    path integral representation of Eq.~\eqref{eq:h}, with non-zero
    $h_z$, can be written down using eigenstates of both $\sigma_x$
    and $\sigma_z$.  Integrating out the bath generates a long-range
    (in time) interaction for $\sigma_x$. Subsequently, the
    $\sigma_z$ degrees of freedom can be integrated out as well,
    leaving a model formulated in $\sigma_x$ only. Re-interpreting the
    $\sigma_x$ values for the individual time slices in terms of Ising
    spins, one arrives at a 1D Ising chain with both short-range and
    $1/r^{1+s}$ interactions, with the thermodynamic limit
    corresponding to the $T\to 0$ limit of the quantum model.
    Repeating this procedure for \twobm\ with $\vec{h}=0$, one obtains
    a Feynman path integral in terms of eigenstates of $\sigma_x$ and
    $\sigma_y$. Importantly, both experience long-range interactions
    and hence neither can be integrated out. This leads to a
    representation in terms of {\em two} coupled Ising
    chains. However, upon re-exponentiating the matrix elements, the
    coupling between the two chains turns out to be imaginary, such
    that a classical interpretation is {\em not} possible
    \cite{tv_pc}. In other words, a Feynman path-integral
    representation of \twobm\ leads to negative Boltzmann weights,
    i.e., a sign problem.

%%%%%%%%%%%%%%%%%%%%%%%%%%%%%%%%%%%%%%%%%%%%%%%%%%%%%%%%%%%%%%%%%%%%%%%

    \emph{Conclusion. ---} Our implementation of OBB-VMPS on the
    Wilson chain brings the Hilbert-space truncation problem of
    the bosonic NRG under control and allows for efficient
    ground-state computations of bosonic impurity models. We have used
    this to verify the QCC in \onebm\ and to determine the phase
    diagram of \twobm, which is shown to violate QCC. This underlines
    that symmetries are decisive
for whether or not   a QCC exists.
A detailed study of the QPTs of \twobm\ is left for future
work.

The results for \twobm\ also show that the predictions from
weak-coupling RG are {\em not} valid for all parameters and bath
exponents, in contrast to expectations. This implies that studying a
{\em three}-bath version of the spin-boson model, which is related to
the physics of a magnetic impurity in a quantum-critical magnet
\cite{rg_bfk,VBS}, is an interesting future subject.

% acknowledgements

We thank A. Alvermann, S. Florens, S. Kirchner, K. Ingersent, Q. Si, A. Schiller
and T. Vojta for helpful discussions.  This research was supported by
the Deutsche Forschungsgemeinschaft through SFB/TR12, SFB631, FOR960, by
the German-Israeli Foundation through G-1035-36.14, and the NSF through
PHY05-51164.

%%%%%%%%%%%%%%%%%%%%%%%%%%%%%%%%%%%%%%%%%%%%%%%%%%%%%%%%%%%%%%%%%%%%%%%

\end{document}

\title{Supplementary Information for ``Critical and strong-coupling phases in one- and two-bath spin-boson models''}

\author{Cheng Guo}
\author{Andreas Weichselbaum}
\author{Jan von Delft}
\affiliation{Ludwig-Maximilians-Universit\"at M\"unchen, Germany}
\author{Matthias Vojta}
\affiliation{Institut f\"ur Theoretische Physik, Technische Universit\"at Dresden, 01062 Dresden, Germany}

\date{\today}

\maketitle

%%%%%%%%%%%%%%%%%%%%%%%%%%%%%%%%%%%%%%%%%%%%%%%%%%%%%%%%%%%%%%%%%%%%%

\section{Discretization and mapping to the Wilson chain}

The spin-boson model represents a prototypical quantum-impurity setup,
with the bath consisting of non-interacting particles. As such it
is amenable to the concept of energy scale separation present in the
NRG \cite{classicNRG,bulla_numerical_2003,bulla_numerical_2005}. For
this, the quantum impurity Hamiltonian of the spin-boson model is
mapped onto a so-called Wilson chain, which includes two steps: (i)
coarse graining of the bath (logarithmic discretization), followed
(ii) by a mapping onto a semi-infinite bosonic chain with the
spin-impurity connected to its starting point.

The bath spectral function $J_{i}(\omega)$ of each bosonic bath $i$
is assumed to be non-zero in the interval $\omega\in[0,\omega_{c}]$,
with $\omega_{c}$ an upper cutoff frequency. The bath spectral
function $J_{i}(\omega)$ is discretized then in energy-space into
intervals $[\omega_{m+1},\omega_{m}]$, marked by the decreasing
sequence $\omega_{m}$ ($m=0,1,\ldots$) with $\omega_{0}=\omega_c$ and
$\lim_{m\to\infty}\omega_{m}=0$. Assuming two identical baths
\mbox{$i\in\{x,y\}$} that couple to the Pauli matrices
$\sigma_{i}$ of the impurity, respectively, the discretized
Hamiltonian has the form
\begin{equation}
   {\cal H}_{\rm  bath} = \sum_{i=x,y} \sum_{m=0}^{\infty}
   \bigl[\xi_{m} \hat{B}_{m i}^{\dagger} \hat{B}_{m i}
   +\frac{\sigma_{i}}{2\sqrt{\pi}}\gamma_{m}(\hat{B}_{m i} + \hat{B}_{m i}^{\dagger})
   \bigr],
\label{eq:Hsb_star}
\end{equation}
Here $\hat{B}_{m i}^{\dagger}$ ($\hat{B}_{m i}$) is the creation
(annihilation) operator, respectively, of a free boson with energy
$\xi_{m}$, that is coupled to the impurity spin with strength
$\gamma_{m}$. Moreover,
\begin{align}
\xi_{m} & =
\frac{\int_{\omega_{m+1}}^{\omega_{m}}J(x)dx}
{\int_{\omega_{m+1}}^{\omega_{m}}\bigl(J(x)/x\bigr)dx} \nonumber \\
\gamma_{m} & =\left(\int_{\omega_{m+1}}^{\omega_{m}}J(x)dx\right)^{1/2}.
\label{eq:Hsb_star_factors}
\end{align}
We prefer a logarithmic discretization scheme over a linear or power-law
discretization, since the study of critical behavior requires
very small energy scales to be resolved. This would require
too many chain sites for linear and even power-law discretization
schemes. Moreover, logarithmic discretization is ideally suited to
represent scale-invariant physics near a quantum phase transition, and it
has the advantage that characteristic NRG information, such as the
energy flow diagram used to analyze the fixed points of the system,
can be extracted from our VMPS results, if desired. (This will be
elaborated upon in a separate publication \cite{Guo2011b}.)

In this paper we use the improved logarithmic discretization recently
proposed by \v{Z}itko and Pruschke \cite{zitko_energy_2009}. As
this achieves a more consistent description of the bath, it reduces
discretization effects and hence allows to determine phase boundaries
such as the critical coupling strength $\alpha_c$ more
accurately. Thus we choose the discretization intervals as
\begin{eqnarray}
   \omega_{0}^{z} & = & \omega_{c},\nonumber \\
   \omega_{m}^{z} & = & \omega_{c}\Lambda^{1-m-z},\quad (m=1,2,3,\ldots)
\label{eq:rok}\end{eqnarray} %
with $\Lambda>1$ Wilson's logarithmic discretization parameter
\cite{classicNRG}, and $z\in\,]0,1]$ an arbitrary shift
\cite{oliveira_1992}. By solving the differential equation in App. C
of Ref.~\onlinecite{zitko_energy_2009} analytically, we obtain
the following explicit expressions
for the parameters in Eq.~(\ref{eq:Hsb_star_factors}):
\begin{eqnarray}
   \xi_{0}^{z} &=& \Bigl[
      \tfrac{1-\Lambda^{-z(1+s)}}{(1+s)\mathrm{ln}\Lambda}-z+1
   \Bigr]^{\tfrac{1}{1+s}}, \nonumber \\
   \xi_{m}^{z} &=& \Bigl[
      \tfrac{\Lambda^{-(1+s)(m+z)}(\Lambda^{1+s}-1)}{(1+s)\mathrm{ln}\Lambda}
   \Bigr]^{\tfrac{1}{1+s}}, \quad (m=1,2,3,\ldots) \nonumber \\
\end{eqnarray}
\begin{eqnarray}
   \gamma_{0}^{z} &=& \sqrt{
      \tfrac{2\pi\alpha}{1+s}(1-\Lambda^{-z(1+s)})},\nonumber \\
   \gamma_{m}^{z} &=& \sqrt{
      \tfrac{2\pi\alpha}{1+s}(\Lambda^{1+s}-1)\Lambda^{-(m+z)(1+s)}}
      \quad (m=1,2,3,\ldots) \nonumber \\
\end{eqnarray}
Having discretized the Hamiltonian, the mapping onto the Wilson chain
is done numerically using standard Lanzcos tridiagonalization. For the
calculations in this paper, we use $z=1$.

\begin{figure}
\includegraphics[width=1\linewidth]{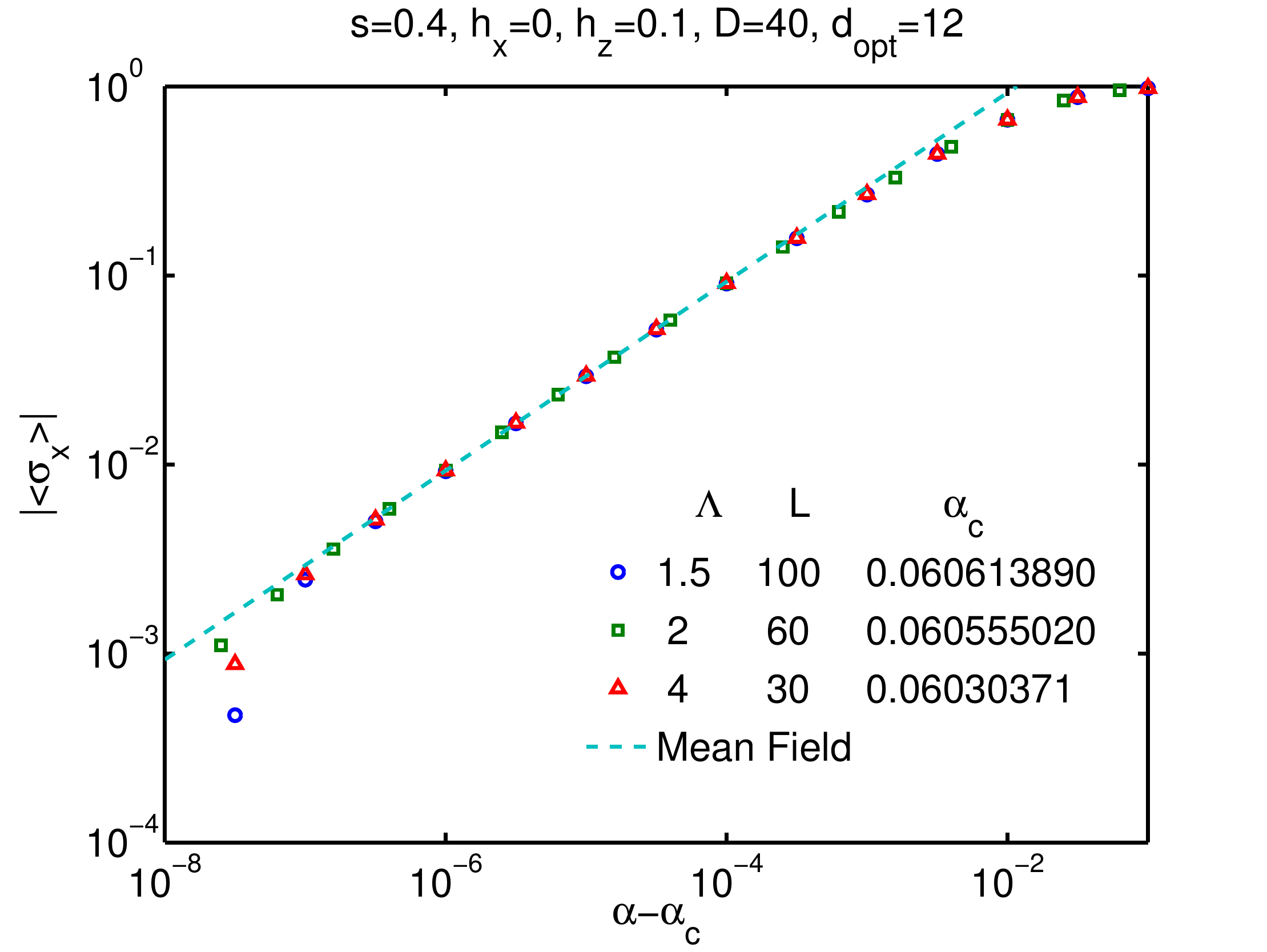}
\caption{
$\langle \sigma_x \rangle$ as function of $(\alpha-\alpha_\critical)$
for \onebm, using
three different choices of the discretization parameter $\Lambda$
(and, correspondingly,
different chain lengths $L$). The data illustrate that the critical
 exponent $\beta$,
obtained from power-law fits to this data, is essentially independent
of $\Lambda$. The
dashed line represents a power law
$\langle \sigma_x \rangle \propto
(\alpha-\alpha_\critical)^\beta$ with
mean-field exponent $\beta_\MF=1/2$.
\label{Flo:fig_convergence_alphac}}
\end{figure}

We find that, with the discretization scheme of \v{Z}itko and Pruschke,
\cite{zitko_energy_2009} the $\Lambda$-dependence of $\alpha_c$ is
much weaker than for the traditional discretization scheme
\cite{classicNRG,bulla_numerical_2005}, i.e., $\alpha_c$ converges
rapidly as $\Lambda$ is decreased towards 1. Similar to standard NRG,
critical exponents do not dependent on $\Lambda$, as shown in
Fig.~\ref{Flo:fig_convergence_alphac} for the exponent $\beta$.

%%%%%%%%%%%%%%%%%%%%%%%%%%%%%%%%%%%%%%%%%%%%%%%%%%%%%%%%%%%%%%%%%%%%%

\section{OBB-VMPS Optimization Procedure}

As discussed in the main paper,
we use a MPS of the following general form:
\begin{equation}
  |G\rangle= \sum_{\spin = \uparrow, \downarrow} \sum_{ \{ \vec n \}}
  A^0[\spin] A^{1}[n_{1}] \cdots A^{L-1}[n_{L-1}] |\spin \rangle |\vec n
  \rangle \; .
\label{eq:mps}
\end{equation}
The $A$-matrix elements are represented as
\begin{equation}
(A^k[n_k])_{\alpha \beta} = \sum_{\tilde
  n_k = 0}^{d_\optimal-1} (\tilde A^k[\tilde n_k])_{\alpha \beta}
V^k_{\tilde n_k n_k} \quad (k \ge 1) \; ,
\label{eq:A_decomposition}
\end{equation}
in order to allow for the construction of an effective
optimized boson basis (OBB) on each site $k$, given by
\begin{equation}
\label{eq:optimizedbosonbasis}
|\tilde
n_k \rangle = \sum_{n_k = 0}^{d_k-1} V^k_{\tilde n_k n_k} |n_k \rangle \;
\quad (\tilde n_k = 0, \ldots ,d_\optimal-1) \; .
\end{equation}
The VMPS ansatz (\ref{eq:mps}) for the ground state of the Wilson
chain is completely analogous to standard finite-size DMRG
\cite{white_density_matrix_1992}, and the use of an optimized local
basis [\Eqs{eq:A_decomposition} and (\ref{eq:optimizedbosonbasis})]
was pioneered in Ref.~\onlinecite{zhang_density_1998}, finding
subsequent applications in, for example,
Refs.~\onlinecite{weie_optimized_2000,nishiyama_numerical_1999}. The
variational optimization of the resulting MPS with respect to the
Hamiltonian ${\cal H}_{\rm sb}^{(L)}$ given in the main paper,
depicted in Fig.~{\figfit} there, proceeds by iteratively updating the
$\tilde A$- and $V$-matrices through a series of sweeps through the
chain.  Given the directed structure of the Wilson chain from large
energy scales (left side of the MPS) to small energy scales (right
side of the MPS), similar to the NRG, variational energy-lowering
updates are performed only when sweeping from left to right. In
contrast, during the reverse sweep from right to left the physical
state (and its energy expectation value) is left unchanged.
Nevertheless, during the reverse sweep the $A$-matrices are recast
into a right-orthonormalized form, to ensure that the right low-energy
part of the Wilson chain is described in terms of properly
orthonormalized \emph{effective} basis sets.

\begin{figure}
\includegraphics[width=1\linewidth]{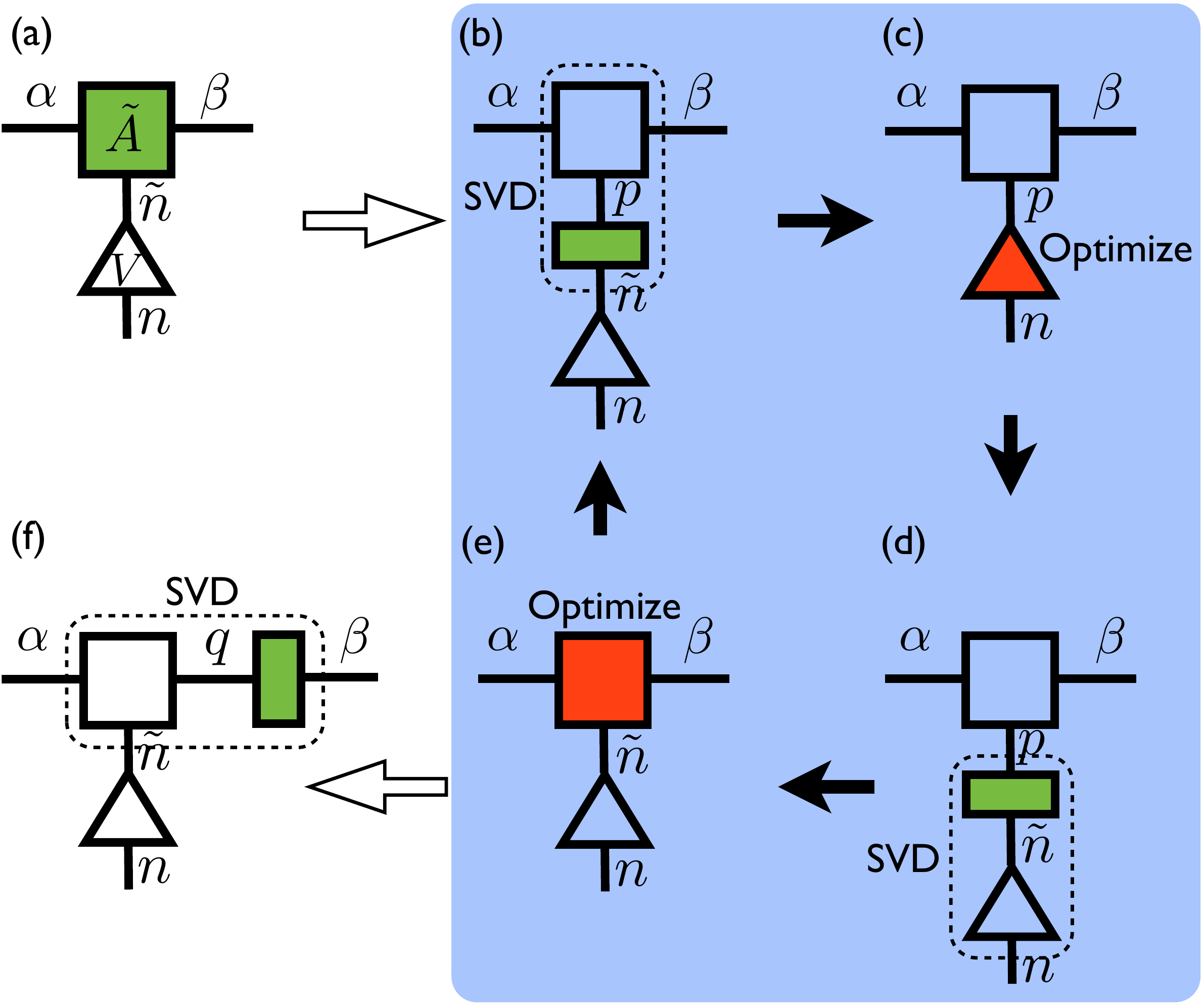}
\caption{Update procedure at one site when sweeping from left to right. 
  The matrices surrounded by the dashed lines are the outcome of
  singular value decomposition (SVD). The filled squares or triangles
  indicate the ``current focus'' of each step.  Filled arrows indicate
  the local update loop (highlighted by the shaded area), open arrows
  indicate its entry and exit.  }
  \label{Flo:fig_mps_method1}
\end{figure}

 To get started, the MPS is
always initialized randomly, followed by proper orthonormalization.
To update the coefficient spaces related to a given site $k$
  of the Wilson chain, we perform the following steps, depicted
  schematically in Fig.~\ref{Flo:fig_mps_method1} (we suppress the
index $k$ below):
\begin{enumerate}
\item[(a)] At a given site, the starting point is a well-defined
  local setting in terms of orthonormal basis sets
  $\vert\alpha\rangle$, $\vert\beta\rangle$, and
  $\vert\tilde{n}\rangle$ of the $A$-tensor for an effective left,
  right, and full local state space, respectively, as depicted in
  Fig.~\ref{Flo:fig_mps_method1}(a). The current approximation to the
  overall ground-state wave function therefore has its focus on the
  $\tilde{A}$-tensor of site $n$, and, setting
    $\tilde{A}_{\alpha\beta\tilde{n}} \equiv (\tilde{A}[\tilde n])_{\alpha
      \beta}$, can be written as
\begin{equation}
   \vert\psi\rangle \equiv \sum_{\alpha\beta\tilde{n}}
   \tilde{A}_{\alpha\beta\tilde{n}} \vert\alpha\rangle \vert\beta\rangle
   \vert\tilde{n}\rangle.
\label{wfA}
\end{equation}
\item[(b)] Downward-orthonormalization of $\tilde A$-tensor to move the focus to
  the $V$-matrix: combine the left state space $\alpha$ with the
  right state space $\beta$ into one index, and use the
  effective local state space $\tilde{n}$ as another index. Then singular
  value decomposition (SVD) of the resulting matrix,
\begin{equation}
    \tilde{A}_{\alpha\beta\tilde{n}} =
    \tilde{A}_{(\alpha\beta)\tilde{n}}=\sum_{p}\tilde{A}'_{(\alpha\beta)p}\lambda_{p}B_{p\tilde{n}} ,
\end{equation}
    generates a new orthonormal tensor $\tilde{A}'$ that describes a
    combined description of the product space $(\alpha,\beta)$, such
    that
\begin{equation}
    \sum_{(\alpha\beta)}\tilde{A}'_{(\alpha\beta)p}\tilde{A}'^{*}_{(\alpha\beta)p'}=
    \delta_{pp'}.
\end{equation}
The wave function can now be represented as
\begin{equation}
    \vert\psi\rangle = \sum_{p \tilde{n}} \lambda_{p}B_{p \tilde{n}}\vert p \rangle \vert\tilde{n} \rangle.
    \label{wfB}
\end{equation}
This description of the rest of the system in terms of an orthonormal
effective basis $\vert p\rangle$ is used to represent operators that
connect the rest of the system with the current local state
space. Being orthonormal, it also makes the numerics simple and
stable.

\item[(c)] By contracting the remaining box $B$ from the last step
  onto the $V$-matrix, this shifts the focus down to the
  $\tilde{V}$-matrix
\begin{equation}
    \tilde{V}_{p n} = \sum_{\tilde{n}} \lambda_{p} B_{p \tilde{n}} V_{\tilde{n} n},
\end{equation}
which thus has been altered.  Having shifted the
focus, the ground state is now represented
as
        \begin{equation}
   \vert\psi\rangle \equiv \sum_{p, n}
   \tilde{V}_{p n} \vert p \rangle \vert n \rangle,
   \label{wf2}
     \end{equation}
     where the $\tilde{V}$-matrices connect the orthonormal state
     spaces $p$ and with the local boson space $n$.  So far, the
     transformation of $|\psi\rangle$ has been exact with
     Eq.~(\ref{wf2}) describing the same state as Eq.~(\ref{wfA}).
   Transform the Hamiltonian and operators into the basis
   $|p\rangle|n\rangle$ and solve the eigenvalue problem
\begin{equation}
\sum_{(p n)}H_{(p'n')(p n)}\tilde{V}_{(p n)}=E_{g}\tilde{V}_{(p' n')},
\end{equation}
for the ground state of the system. Thus
the resulting $\tilde{V}$-matrix represents the \emph{updated}
ground state via \Eq{wf2}.

\item[(d)] Upward-orthonormalization of the $\tilde{V}$-matrix:
    the singular value decomposition of $\tilde{V}$,
\begin{equation}
    \tilde{V}_{n p}=\sum_{\tilde{n}}V_{n \tilde{n}}s_{\tilde{n}}C_{\tilde{n} p} \; ,
\label{eq:svd_V}
\end{equation}%
provides a new effective description of the local boson space,
such that the transformation matrix $V$ from the
original boson basis to the OBB
is orthogonal:
\begin{equation}
  \sum_{n}V_{n\tilde{n}}V_{n\tilde{n}'}^{*}=\delta_{\tilde{n}\tilde{n}'}.\end{equation}
Note that the singular values $s_{\tilde{n}}$ indicate the relative importance
of the optimal boson bases. We will come back to this point at the
end of this section. At this step the wave function is
\begin{equation}
    \vert\psi\rangle = \sum_{\tilde{n} p} s_{\tilde{n}}C_{\tilde{n} p}\vert\tilde{n} \rangle \vert p \rangle .
    \label{wfC}
\end{equation}

\item[(e)] By contracting the remaining
box $s_{\tilde{n}}C$ from the previous step onto the $\tilde{A}'$-matrix,
similar with step (c),
the focus can be shifted to the $\tilde{A}$-matrix again:
\begin{equation}
    \tilde{A}_{\alpha \beta \tilde{n}} = \sum_{p} \tilde{A}'_{\alpha \beta p} s_{\tilde{n}}C_{\tilde{n} p}.
\end{equation}
The wave function is thus reexpressed in the same form as
  Eq.~(\ref{wfB}).  Now transform the local operators to the OBB
$|\tilde{n}\rangle$ using $V$, and optimize the
$\tilde{A}_{\alpha\beta\tilde{n}}$ matrix in the same way as done
using the traditional VMPS method.
\item[(f)] Combine the left and local indices and perform a singular value
  decomposition of the $\tilde{A}_{\alpha\beta\tilde{n}}$
  matrix:
\begin{equation}
    \tilde{A}_{\alpha\beta\tilde{n}}=\tilde{A}_{(\alpha\tilde{n})\beta}=\sum_{q}\tilde{A}_{(\alpha\tilde{n})q}r_{q}F_{q\beta}.\label{eq:svd_A}
\end{equation}
The resulting tensor $\tilde{A}_{\alpha q\tilde{n}}$ is left orthogonal:
\begin{equation}
\sum_{\alpha\tilde{n}}\tilde{A}_{\alpha q\tilde{n}}\tilde{A}^{*}_{\alpha q'\tilde{n}}=\delta_{q q'}.
\end{equation}
Contracting the remaining $r_{q}F_{q\beta}$ to the $\tilde{A}$ tensor on the
right side completes the update of the current site.
\end{enumerate}
The OBB method enables us to increase the number $d_k$ of local states
that can be kept in the original basis from a few dozen to $d_k
\lesssim 10^{4}$.  (In the next section we shall show that by
implementing explicit oscillator shifts, the \emph{effective} number
of local boson states that are accounted for in the unshifted basis
can be increased to more than $10^{10}$.)

\begin{figure}
\includegraphics[width=0.85\linewidth]{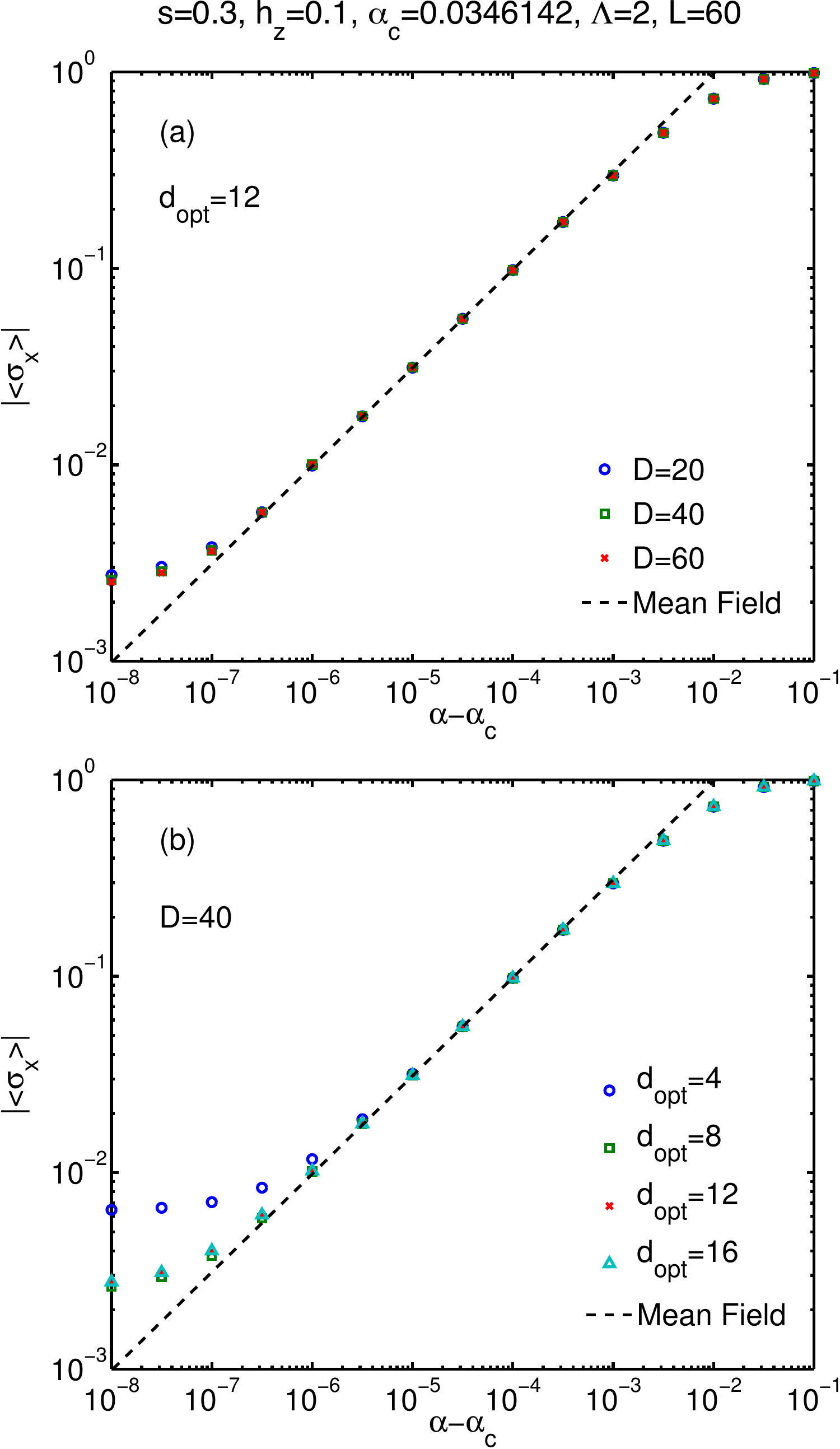}
\caption{Convergence check for the VMPS parameters $D$ (a) and
  $d_\optimal$ (b), for \onebm. The panels show
  $\langle\sigma_{x}\rangle$ as function of $\alpha_x$ for different
  choices of $D$ and $d_\optimal$, respectively.  Since critical
  exponents are obtained from fits to the linear parts of such curves,
  the resulting exponents are evidently not sensitive to $D$ and
  $d_\optimal$. As in Fig.~\ref{Flo:fig_convergence_alphac},
  the dashed lines show the
  mean-field power law with $\beta_\MF=1/2$.
\label{Flo:fig_D_dopt_SVD}}
\end{figure}

The two adjustable VMPS parameters are the dimension $D$ of the VMPS
matrices (corresponding to the number of DMRG states kept) and the
dimension $d_\optimal$ of the optimal boson basis.  To exemplify the
influence of $D$ and $d_\optimal$ on physical quantities,
Fig.~\ref{Flo:fig_D_dopt_SVD} plots the magnetization
$\langle\sigma_{x}\rangle$ for \onebm{} as a function of $\alpha$ for
different $D$ and $d_\optimal$.  Clearly, $\alpha_{\rm c}$ is already
well converged throughout. In practice, we chose $D$ and
    $d_\optimal$ large enough to ensure that all singular values
    [$s_{\tilde{n}}$ and $r_{q}$ in Eqs.~\eqref{eq:svd_V} and
    (\ref{eq:svd_A})] larger than $10^{-5}$ were retained throughout
    the entire Wilson chain, except possibly at  its very end.
    We will explain this in more detail in the next section, when
    discussing Fig.~\ref{Flo:SBM1sv} below.
%More specifically, in our calculation we
%keep a record of $s_{\tilde{n}}$ and $r_{q}$ on all boson sites,
%and we choose $D$ and $d_\optimal$ to make sure that the
%$s_{\tilde{n}}$ and $r_{q}$ on all sites decay down to at least
%$10^{-5}$

For \twobm{} with two bosonic baths, on the other hand, the
combination of two boson sites into one supersite requires numerical parameters
such as $D$ and $d_\optimal$ to be set to larger values than for
\onebm. Nevertheless, we find that the number of kept states
needed to ensure an accuracy comparable to that of \onebm{} is
smaller than the $D^{2}$ or $d_\optimal^{2}$ that might have
been naively expected from the fact that the local state space
now has a direct product structure.

\begin{figure}
\includegraphics[width=0.9\linewidth]{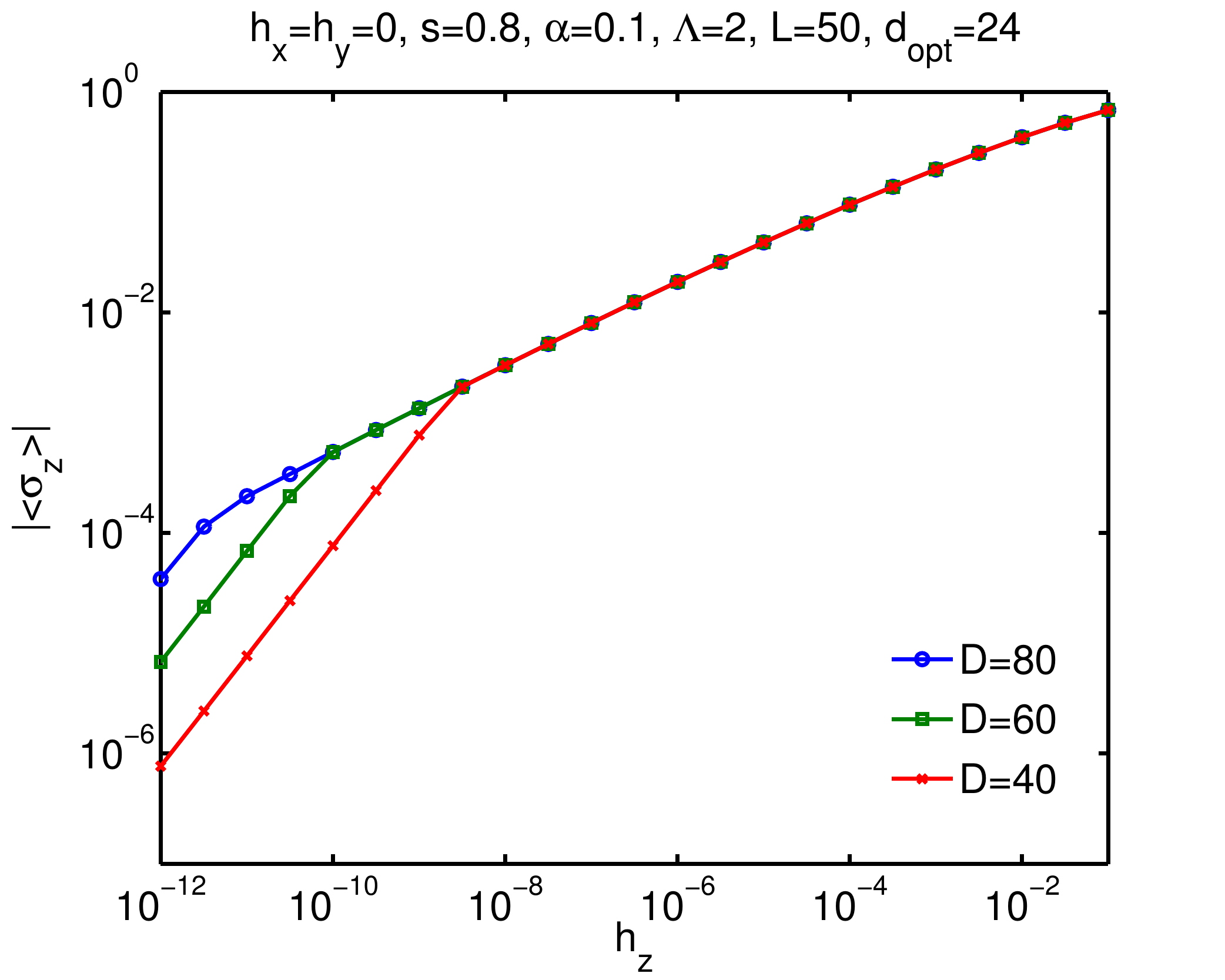}

\includegraphics[width=0.9\linewidth]{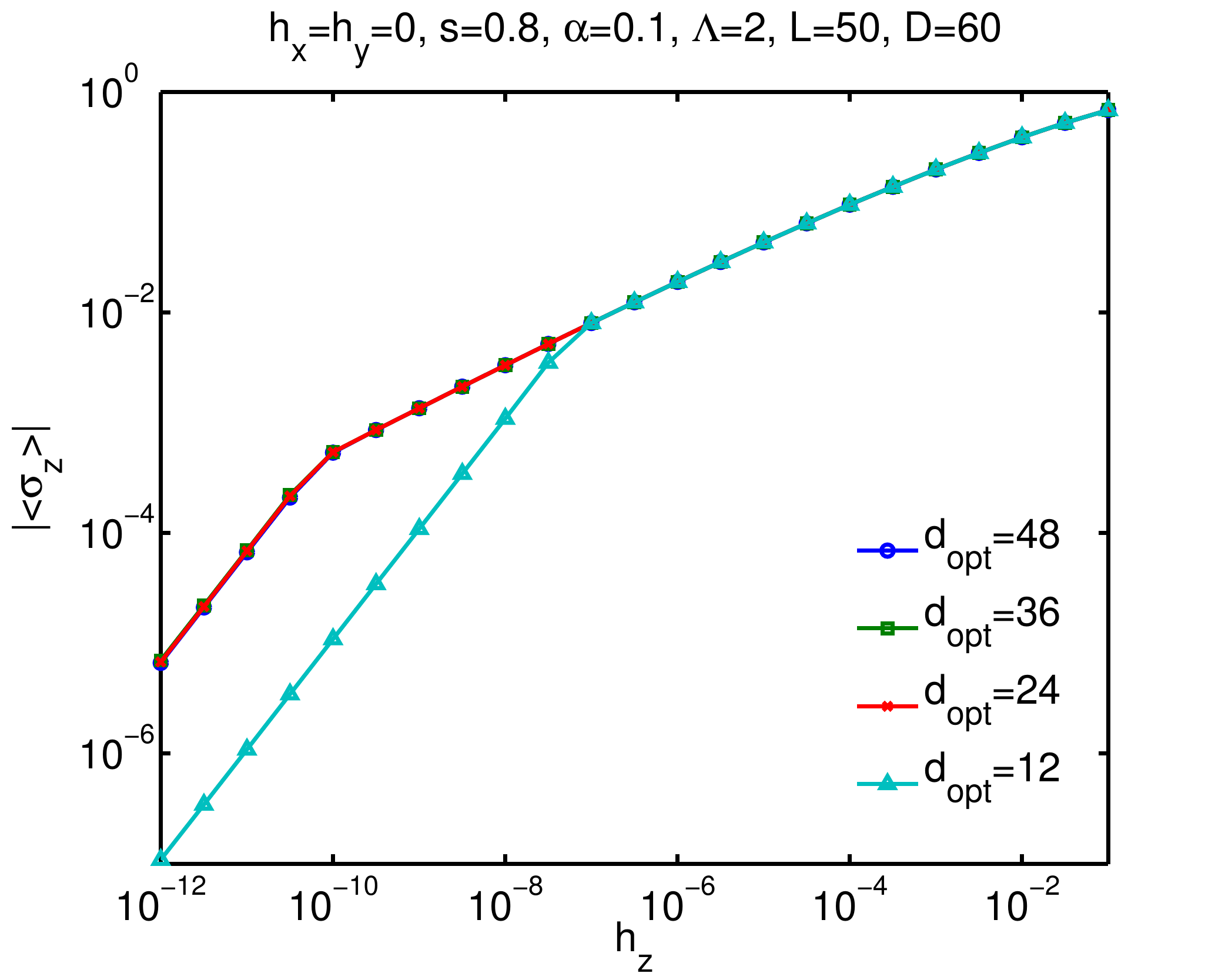}
\caption{
  Convergence check for the VMPS parameters $D$ (a) and $d_\optimal$ (b), for
  \twobm. The panels show the transverse-field response,
    $|\langle \sigma_z \rangle |$ vs.\ $h_z$ in the CR phase, with a
    robust sublinear power law. The deviations at smallest fields
    arise from small symmetry-breaking effects in the numerics which
    cause a linear response akin to an ordered state.
\label{Flo:fig_SBM2_conv}}
\end{figure}

%%%%%%%%%%%%%%%%%%%%%%%%%%%%%%%%%%%%%%%%%%%%%%%%%%%%%%%%%%%%%%%%%%%%%

\section{OBB with explicit shifts}

The OBB scheme described in the preceding section allows us to
  easily perform calculations on a desktop computer using local boson bases of
  dimension $d_k \lesssim 10^{4}$.  This can be increased by at least
  6 more orders of magnitude by analytically incorporating explicit
  shifts during the construction of OBB. The idea is to explicitly
  shift the harmonic oscillator coordinates $\hat{x}_k$ by their
  equilibrium expectation values $\langle \hat{x}_k \rangle$ (similar
  in spirit to the procedure used in Ref.~\onlinecite{fehske}), such that
  the OBB can be used to capture the quantum fluctuations of the
  shifted coordinate
\begin{eqnarray}
\label{eq:desiredshift}
\hat{x}'_k = \hat{x}_k - \langle \hat{x}_k \rangle \; .
\end{eqnarray}
 We now describe explicitly how this is done.

We begin by noting that a shift corresponds to a unitary transformation
\begin{equation}
   \hat{U}(\delta_{k})% \equiv e^{-i\delta_{k} \hat{p}_{k}}
 = e^{\tfrac{\delta_{k}}{\sqrt{2}}(\hat{b}_{k}^{\dagger}-\hat{b}^\pdag_{k})}
\label{eq:shift-transformation}
\end{equation}
such that,
\begin{eqnarray}
   \hat{b}'_{k} &=& \hat{U}^{\dagger}(\delta_{k})
   \,\hat{b}_{k}\, \hat{U}(\delta_{k})
 = \hat{b}_{k}+\frac{\delta_{k}}{\sqrt{2}}
\label{eq:shiftb}
\end{eqnarray}
on $\hat{b}_k$ (similarly for $\hat{b}_k^\dagger$). Thus the
  harmonic oscillator displacement $\hat
  {x}_{k}\equiv\frac{1}{\sqrt{2}}(\hat
  {b}_{k}^\pdag+\hat{b}^\dagger_{k})$ is shifted to $\hat{x}_k' =
  \hat{x}_k + \delta_k$, and the local boson number operator to
\begin{equation}
  \hat{n}'_{k} \equiv \hat{b}_{k}^{\prime \dagger}
\hat{b}_{k}^{\prime \phantom\dagger}
  = \hat{n}_{k} + \delta_{k} \hat{x}_{k} + \tfrac{\delta_{k}^{2}}{2}
  \text{.}\label{eq:shiftn}
\end{equation}
The shift can be implemented on the Hamiltonian level,
by replacing the original Wilson-chain Hamiltonian by
the shifted Hamiltonian
\begin{eqnarray}
\label{eq:shift-H}
{\cal H}_{\rm sb}^{\prime (L)} (\hat{b}_{k i}, \hat{b}_{k i}^{\dagger}) =
{\cal H}_{\rm sb}^{(L)} (\hat{b}_{k i}', \hat{b}_{k i}^{\prime \dagger}) \; .
\end{eqnarray}
The local states $|n_{k}\rangle$ in Eq.~(\ref{eq:optimizedbosonbasis})
now represent Fock states of the shifted oscillators.

To incorporate explicit shifts within the OBB sweeping strategy
described in the last section, we calculate the ground state
expectation value $\langle \hat{x}_k \rangle$ of the current site
after step (e). We then set $\delta_k = - \langle \hat{x}_k \rangle$
in Eqs. \eqref{eq:shiftb} and (\ref{eq:shiftn}), thereby ensuring that
the shifted coordinate $\hat{x}_k'$ corresponds to
\Eq{eq:desiredshift}. We subsequently move back to step (b) and
implement the shift in the Hamiltonian. (In practice, it is convenient
to preserve the form of the Hamiltonian
itself, and instead to change the matrix representation of the operators
$\hat b_k$, $\hat b_k^\dagger$ and $\hat n_k$
to implement the shift of Eq. (24).) Then we repeat the local update loop of
the sequence (b), (c), (d), (e) (Fig.~\ref{Flo:fig_mps_method1}) until
the shift converges.  Finally, we move to step (f) and the next site.

\begin{figure}
\includegraphics[width=0.9\linewidth]{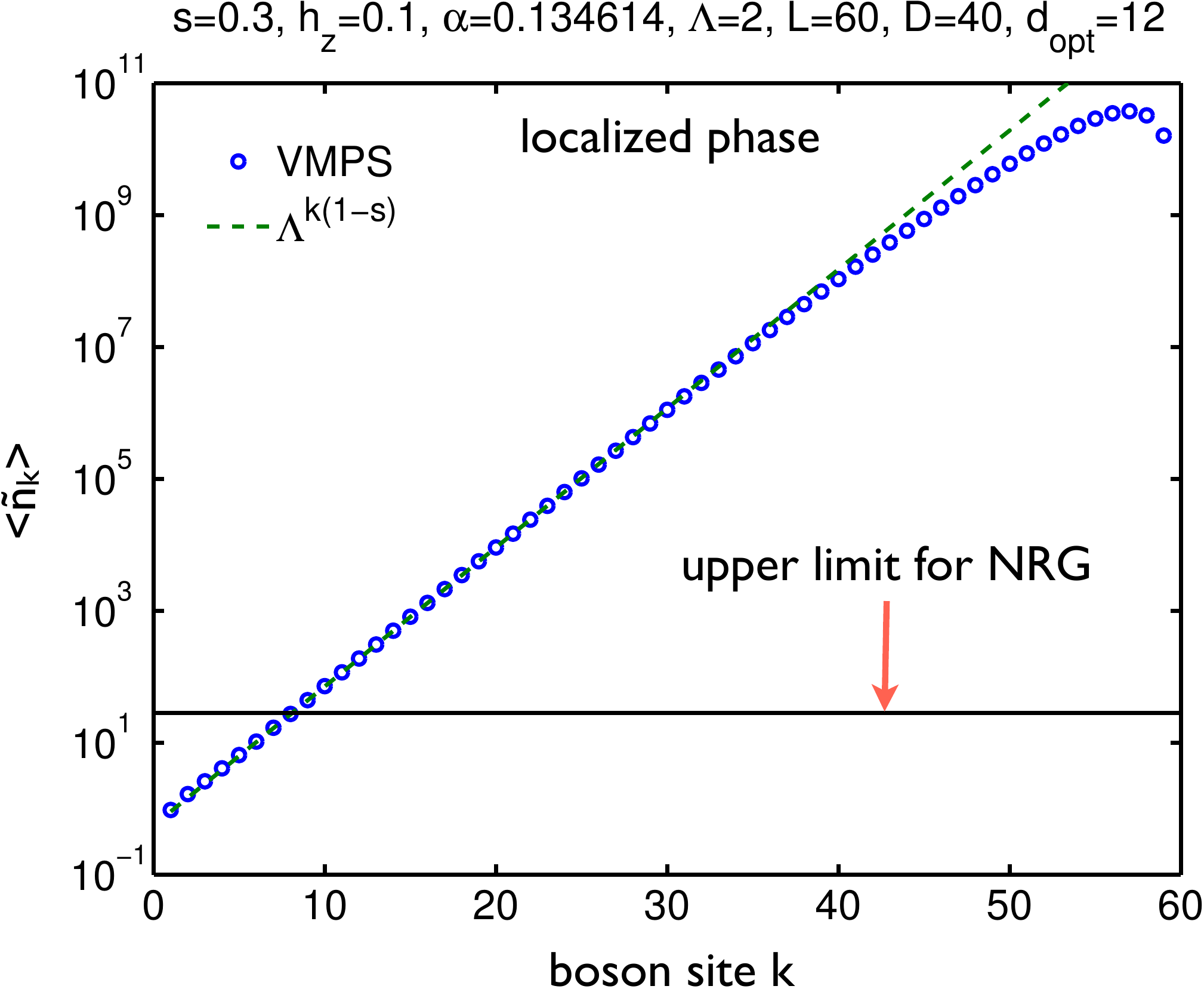}
\caption{Effective boson occupation number $\langle \hat{n}_k \rangle$
  in the original, unshifted Wilson-chain boson basis, as
  function of site index $k$ along the Wilson chain, calculated for
  the localized phase of \onebm.  Dashed line indicates the relation
  $\langle \hat{n}_k \rangle \propto \Lambda^{k (1-s)}$ (see text).
  The finite size effect causes VMPS result deviate from this
  exponential relation at the end of the Wilson chain.}
\label{Flo:nk}
\end{figure}

The SBB method allows us to reach boson shifts $\langle \hat x_k
\rangle$ so large that their description within the unshifted boson
basis would require local dimensions of order $d^\eff_k \simeq
10^{10}$, while nevertheless keeping the actual number of boson states
in the \emph{shifted} basis reasonably small, typically $d_k \lesssim
10^2$.  A typical result for the resulting boson occupation numbers
$\langle \hat{n}_k \rangle$ in the original, unshifted Wilson-chain
boson basis is shown in Fig.~\ref{Flo:nk}, calculated in the localized
phase.  Since the boson shifts for the bosons in the original
definition of \onebm\ scale as\cite{bulla_numerical_2005} $\langle \hat x_k \rangle \simeq
\omega_k^{(s-1)/2}$, with $\omega_k \propto \Lambda^{-k}$, we expect
and indeed find that $\langle \hat{n}_k \rangle$ increases
exponentially with $k$, as $\Lambda^{k (1-s)}$.

\begin{figure}
\includegraphics[width=0.9\linewidth]{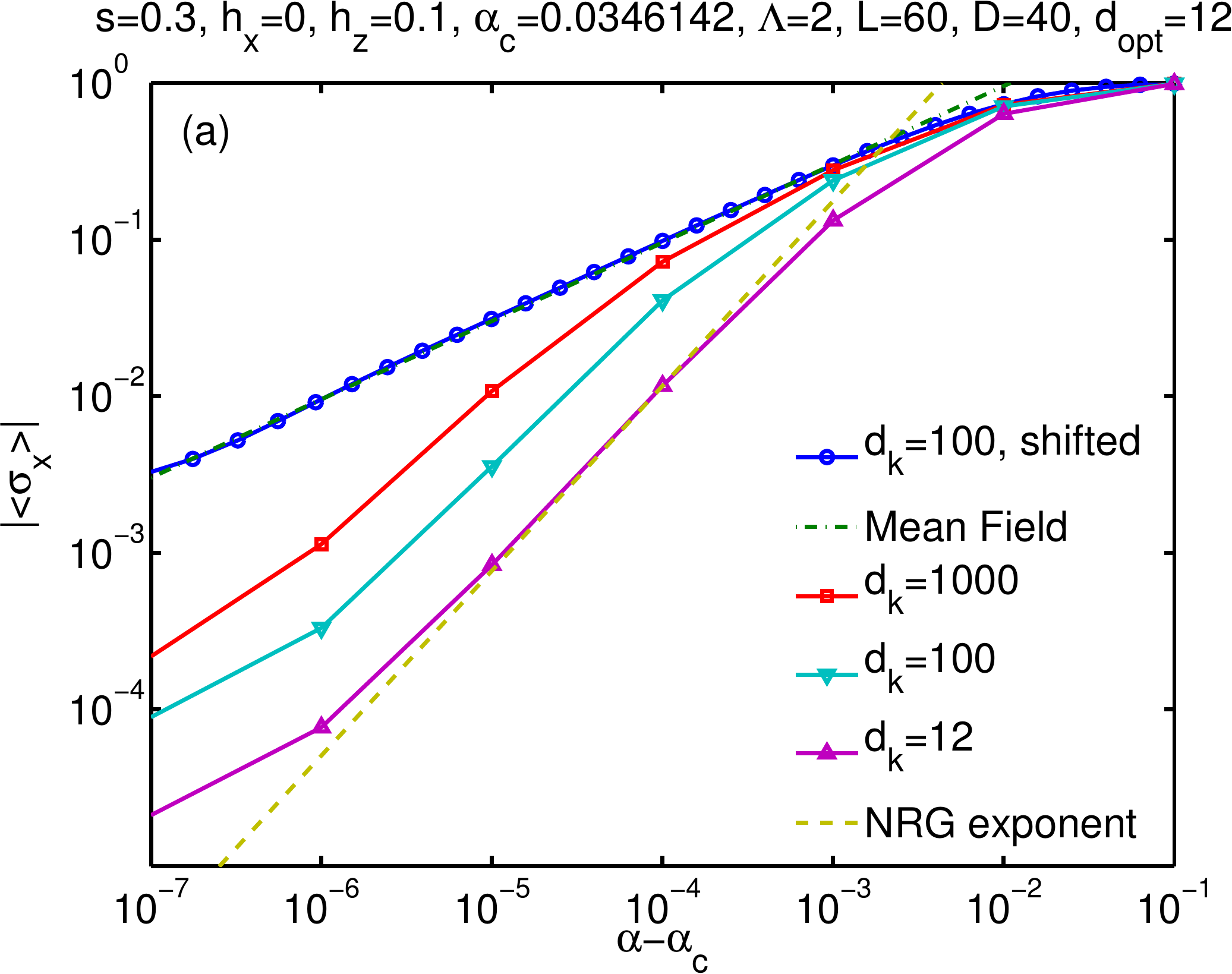}
\includegraphics[width=0.9\linewidth]{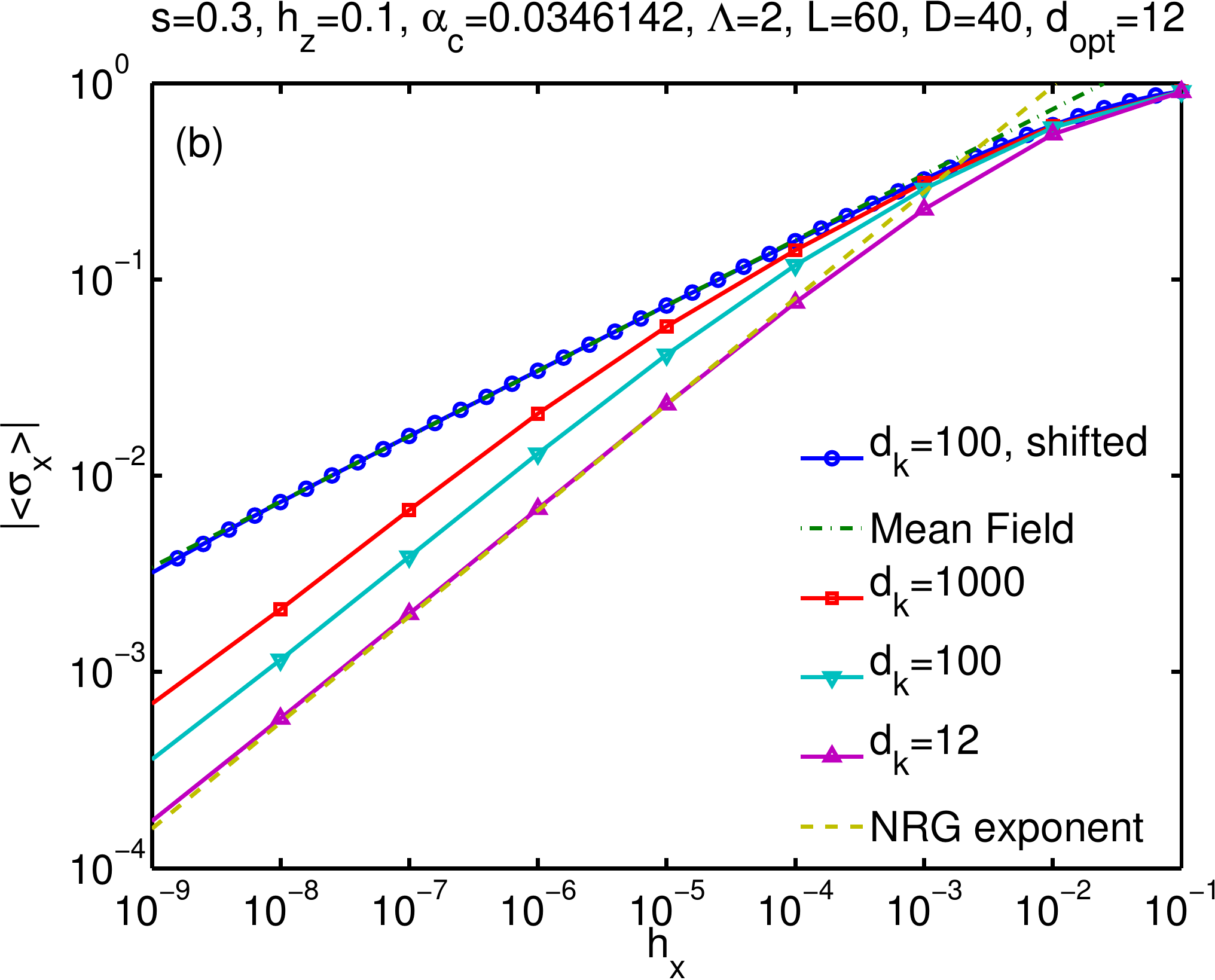}
\caption{The effect of Hilbert space truncation on the
  calculation of the \onebm\ critical exponents $\beta$ in (a) and
  $\delta$ in (b).  The calculation labeled ``$d_k = 100$, shifted''
was performed using the OBB method with explicit shifts, without
setting any maximum value for the size of the shifts,
i.e.\ without restricting $d^\eff_{k}$.
Dashed straight lines correspond to power laws with exponents $\beta_{\rm NRG}=
1.18$ and $\delta_{\rm NRG}= 1.85$ obtained by NRG \cite{VTB}.
$\delta_{\rm NRG}$ agrees
with the hyperscaling exponent $\delta = (1+s)/(1-s)$
that applies for $s>1/2$, see Sec.~\ref{sec:hyper} below.\cite{mv12}
\label{Flo:betafitdiffdkmax}}
\end{figure}

An accurate representation of this exponential rise, as achieved by
  our VMPS scheme, is essential for obtaining correct results for
  critical exponents. The detrimental effects of Hilbert space
  truncation are illustrated vividly in
  Fig.~\ref{Flo:betafitdiffdkmax}. It shows $\langle \sigma_{x}
  \rangle$ vs.\ $(\alpha-\alpha_c)$ for \onebm, calculated for several
  upper limits on the size $d_k$ of the local boson basis. While the
  exponents obtained by NRG \cite{VTB} (indicated by
  dashed lines) correspond to $d_k \le 12$, the curves clearly change
  strongly as $d_k$ is increased. Indeed, fully converged results are
  obtained only when the local boson basis can be taken to have
  ``unlimited'' size, as is the case for the explicit shifting
  strategy discussed above.

  With the SBB method, the results for chain length $L=60$ shown in this paper
  can be obtained within a few hours on a desktop computer. Note,
  though, that in the localized phase the total time needed for the
  calculation increases exponentially with $L$. The reason is that the
  converged value for the effective shift $\langle \hat{x}_k \rangle$
  increases exponentially with $k$ as
\begin{equation}
  \langle \hat{x}_k \rangle \sim \Lambda^{k (1-s)/2}
  \text{,}\label{eq:expshift}
\end{equation}
as explained above. However, the ``sweeping step size'' for $\langle
\hat{x}_k \rangle$, i.e.\ the change in this quantity from one sweep
to the next, is limited, in effect, by the dimension $d_k$ of the
  shifted local boson basis.  Consequently, the number of sweeps
needed to achieve convergence for $\langle \hat{x}_k \rangle$
increases exponentially with $k$ (and making an informed initial guess
for the requisite shift $\langle \hat{x}_k \rangle$ does not really
help in speeding up its accurate determination). 

\begin{figure}
\includegraphics[width=1\linewidth]{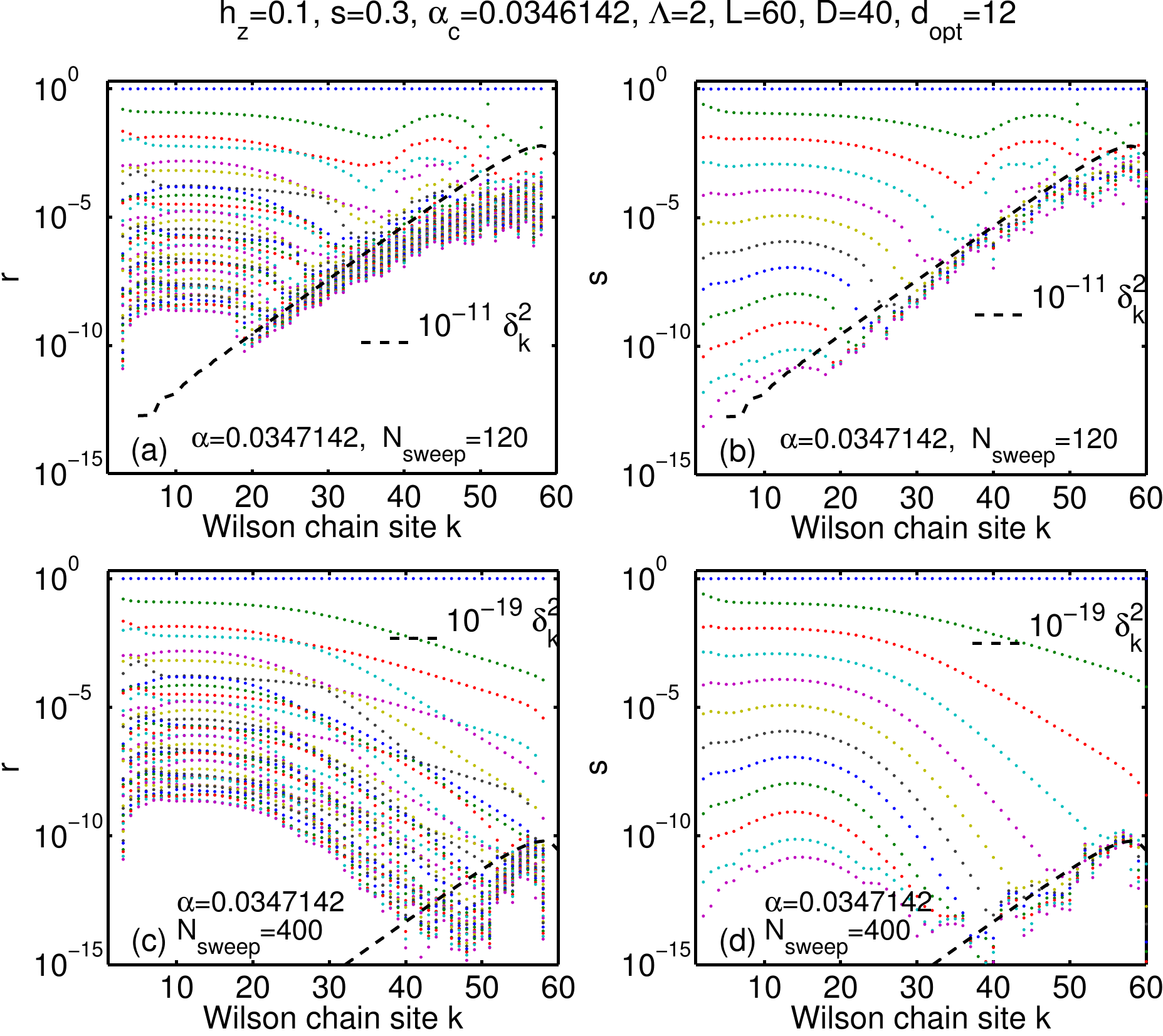}
\caption{
  Panels (a,c) and (b,d) show examples of the singular values
    $r_{q}$ [Eq. \eqref{eq:svd_A}] and $s_{\tilde{n}}$
    [Eq. \eqref{eq:svd_V}], respectively, as functions of the Wilson
    chain $k$, calculated in the localized phase of SBM1.  Panels
    (c,d) show the same quantities as (a,b), but calculated using
    stricter convergence criteria requiring more sweeps, resulting in
    better-converged singular values [see discussion after
    \Eq{eq:expshift}]. The maximum value of $\langle \hat{n}_{k}
    \rangle$ in this example is of order $10^{8}$, corresponding to a
    maximum shift $\langle \delta_{k} \rangle$ is of order $10^{4}$.
    The diagonal dashed lines show the $k$-dependence of the
      shifts $\delta_k^2$, multiplied by a constant prefactor that was
      chosen by hand in such a way that the dashed lines lie near the
      onset of noisiness at the lower end of the singular value
      spectra. This reveals the direct correlation between the
      exponential increase of $\delta_k^2$, which directly enters the
      Hamiltonian, and the noise in the smallest singular values.}
\label{Flo:SBM1sv}
\end{figure}

For a given set of convergence criteria, the exponential
  growth in the shifts $\langle \hat{x}_k \rangle$ is accompanied by a
  similar growth in the absolute errors in their numerical
  determination. The consequences of this for SBM1 can be seen in
  Figs.~\ref{Flo:SBM1sv}(a) and \ref{Flo:SBM1sv}(b), which show
  examples of the singular values $r_{q}$ [Eq. \eqref{eq:svd_A}] and
  $s_{\tilde{n}}$ [Eq. \eqref{eq:svd_V}], respectively, as functions
  of the Wilson chain $k$: when $k$ becomes large, the the lower end
  of the singular value shows an increasing amount of scatter.  This
  implies that these singular values are not yet properly converged,
  which is directly correlated to the uncertainties in the oscillator
  shifts.  Better convergence \emph{can} be achieved by using stricter
  convergence criteria, as illustrated by Figs.~\ref{Flo:SBM1sv}(c)
  and \ref{Flo:SBM1sv}(d), but only at a considerable increase in
  computation time, essentially using up to several hundreds of
  sweeps. We have thus adopted a compromise between accuracy and
  computation time: we chose $D$, $d_\optimal$ and the shift
  convergence criteria such that throughout the entire Wilson chain,
  all discarded singular values were smaller than $10^{-5}$, except
  possibly for an increase in this tolerance at the very end of the
  chain, of the type seen in Figs.~\ref{Flo:SBM1sv}(a) and
  \ref{Flo:SBM1sv}(b). We have checked explicitly that
  not-optimally-converged shifts and singular values towards the end
  of the Wilson chain do not noticably affect the resulting physical
  quantities of interest, i.e. that these are already well converged
  nevertheless.

\begin{figure}
\includegraphics[width=1\linewidth]{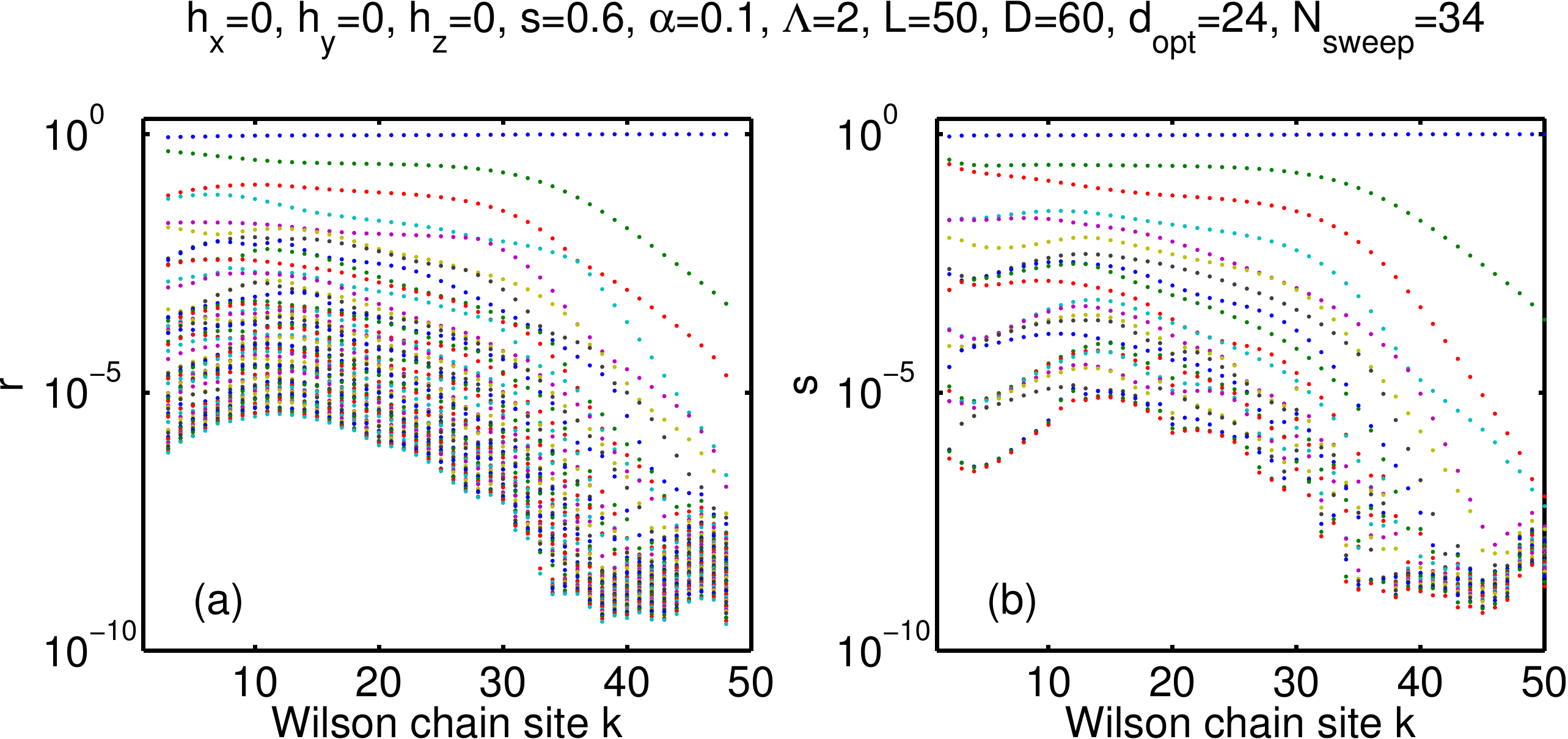}
\caption{Panels (a,b) show examples of the singular
    values $r_{q}$ [\Eq{eq:svd_A}] and $s_{\tilde{n}}$
    [\Eq{eq:svd_V}], respectively, as functions of the Wilson chain
    $k$, calculated in the localized regime of SBM2.  For all the SBM2
    parameters studied in this paper, the maximum effective boson
    occupation number $\langle \hat{n}_{k} \rangle$ is relatively much
    smaller than those we found in SBM1 ($\langle \hat{n}_{k}
    \rangle<10^{5}$), which results a faster convergence of shift
    $\langle \delta_{k} \rangle$.  In this typical example for SBM2,
    we get converged singular values already with $N_{\rm
      sweep}=34$. }
\label{Flo:SBM2sv}
\end{figure}

Interestingly, we have found that for SBM2 it is easier to
  obtain a well-converged shift in the localized phase than for SBM1,
  because the competition between the two chains causes the increase
  in oscillator shifts near the end of the Wilson chain to be much
  smaller for SBM2 than SBM1, as illustrated in Fig.~\ref{Flo:SBM2sv}.
  Near the chain's beginning, in contrast, the singular values for
  SBM2 were found to be bigger than for SBM1, but we nevertheless
  ensured throughout that only singular values below the $10^{-5}$
  tolerance were discarded.

%%%%%%%%%%%%%%%%%%%%%%%%%%%%%%%%%%%%%%%%%%%%%%%%%%%%%%%%%%%%%%%%%%%%%

\section{Additional results for \onebm}

\subsection{Determination of phase boundary}

We have explored a number of different methods
for determining the phase boundary between the localized and delocalized phases,
all of which yield essentially equivalent results. (Fortunately,
no oscillator-shift-related  convergence problems occur in this
context, since the oscillator shifts are essentially zero at the 
phase boundary.)

\emph{1. ``Best power law''.}  As mentioned in the main paper, our
``standard method'' for determining $\alpha_c$ has been to tune it
such that $\langle \sigma_x \rangle$ vs. $(\alpha_x - \alpha_c)$
yields the best straight line on a log-log plot.  This turned out to
be the most convenient way of getting accurate critical exponents.

\emph{2. Energy flow diagrams.} In NRG, energy-flow diagrams
\cite{classicNRG} can be used to accurately determine the phase
boundary of a quantum phase transition, since the degeneracies of
  low-lying levels typically differ for the two phases to be
  distinguished. It is possible to generate such energy-flow
diagrams also within the VMPS approach \cite{Saberi2008}, by
calculating the eigenstates of the left block's Hamiltonian at each
site when sweeping from left (large energy scales) to right (small
energy scales), and then appropriately rescaling and shifting the
  resulting eigenenergies \cite{classicNRG}.  Details of this
  procedure and explicit examples of flow diagrams obtained in this
  manner will be presented elsewhere \cite{Guo2011b}.

Energy-flow diagrams, however, are sensitive to relevant
perturbations in terms of small numerical inaccuracies at large
energies, i.e.\ early Wilson shells. This can lead to the artificial
breaking of symmetries. For example, the Hamiltonian
of \onebm\ at $h_x = 0$ commutes with the parity operator
\begin{equation}
  \hat{P} \equiv \sigma_{x} e^{i\pi\hat{N}}
\text{,}
\end{equation}
where $\hat{N}=\sum_{k} \hat{b}_{k}^{\dagger}
\hat{b}_{k}^{\phantom\dagger}$ counts the total number of bosons on
the entire Wilson chain. The corresponding parity symmetry guarantees
that the ground state is two-fold degenerate. However,
this degeneracy is typically broken by numerical inaccuracies,
unless the symmetry is explicitly implemented in the numerical
code. We have done so, and flow-diagrams resulting from
a parity-symmetric version of our code yield $\alpha_c$ values
in agreement with the ``standard method'' used in the main paper.

\emph{3. Diverging boson number.} Another procedure
for determining $\alpha_c$ is to plot
the boson occupation number  for all sites
of the Wilson chain: a divergence of
$\langle \hat{n}_k \rangle$ with Wilson-chain index $k$,
as seen in Fig.~\ref{Flo:nk}, is a signature of the
localized phase. 

%%%%%%%%%%%%%%%%%%%%%%%%%%%%%%%%%%%%%%%%%%%%%%%%%%%%%%%%%%%%%%%%%%%%%

\subsection{Critical exponents}

\begin{figure}
\includegraphics[width=0.49\linewidth]{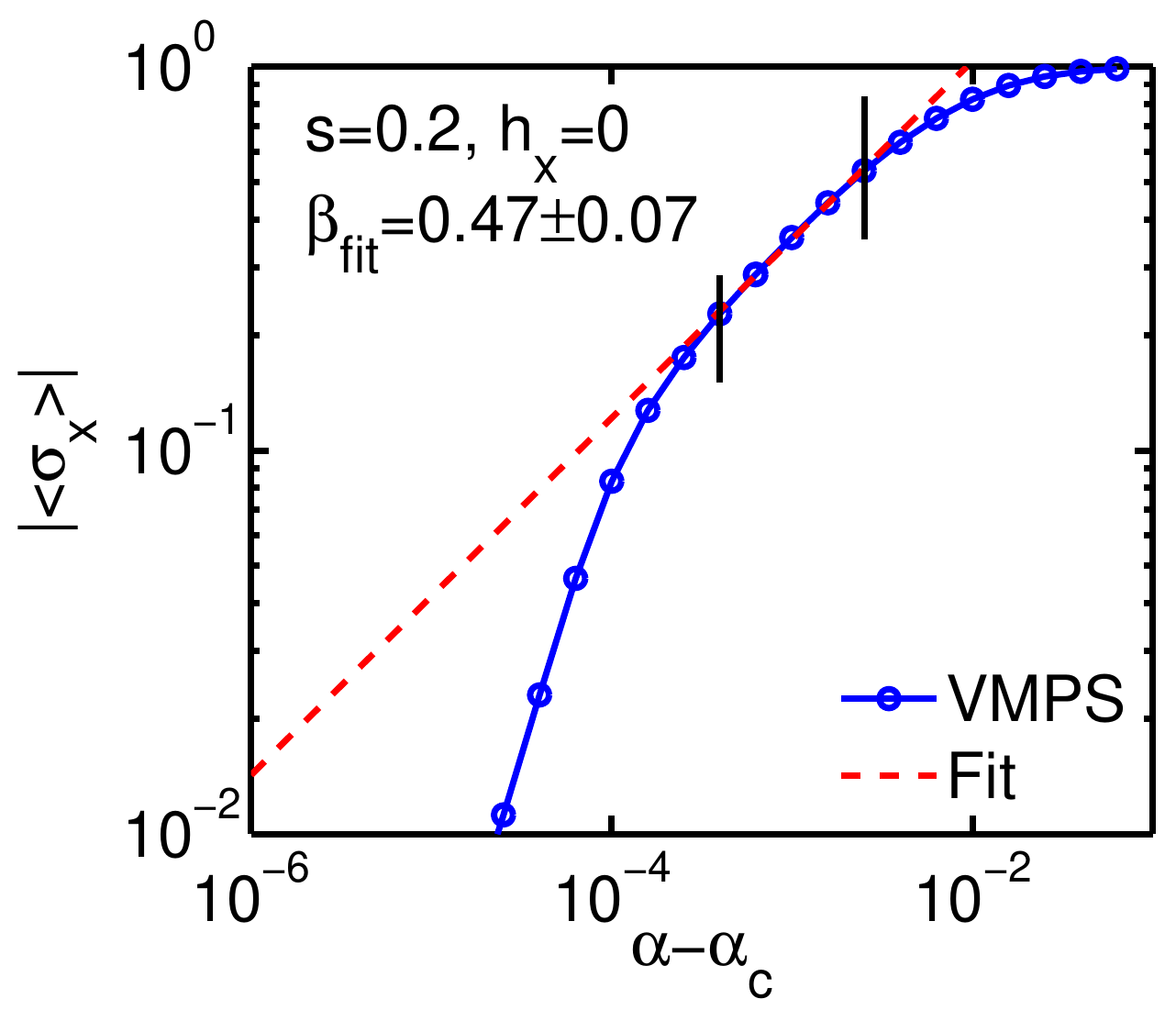}
\includegraphics[width=0.49\linewidth]{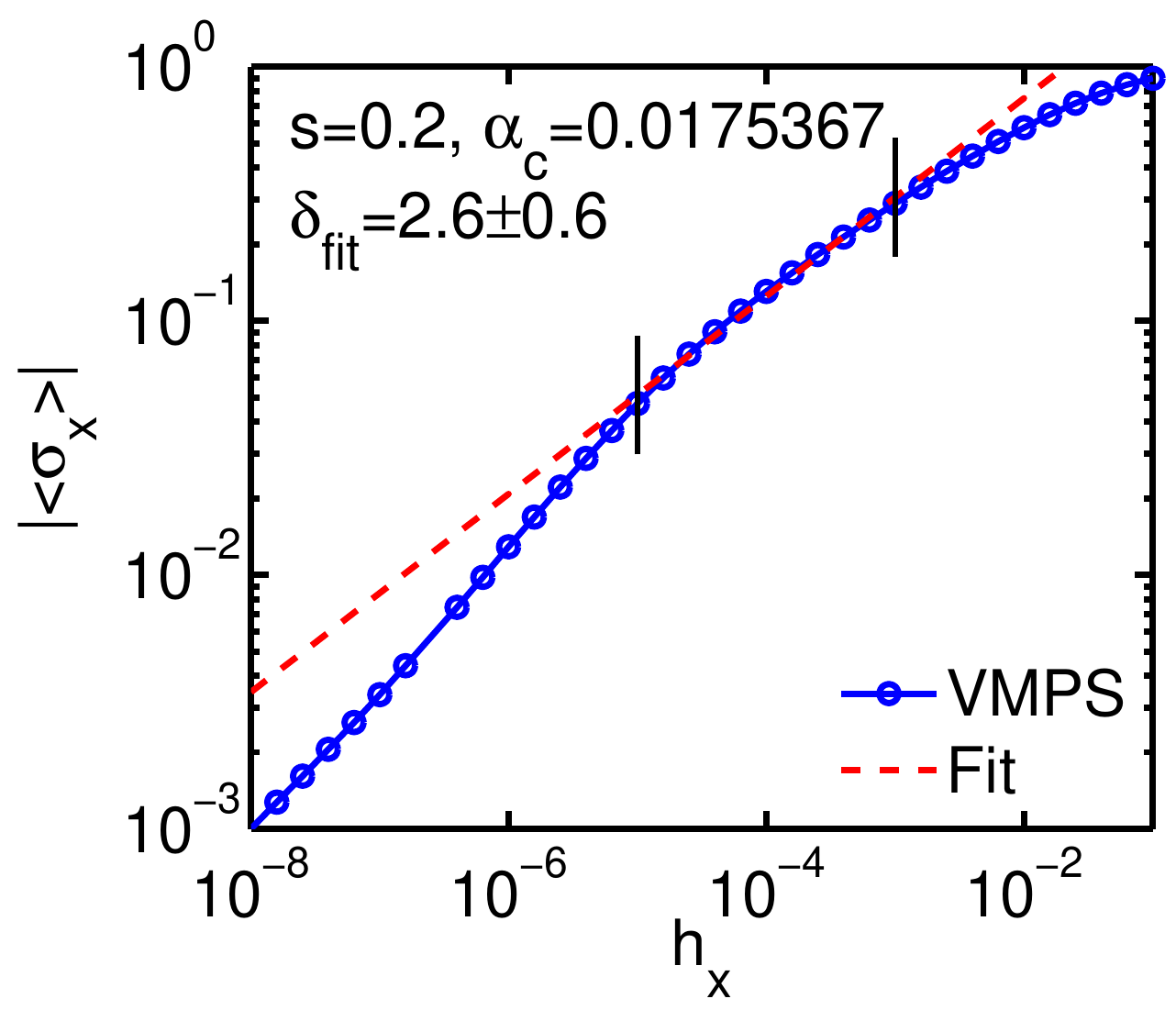}
\includegraphics[width=0.49\linewidth]{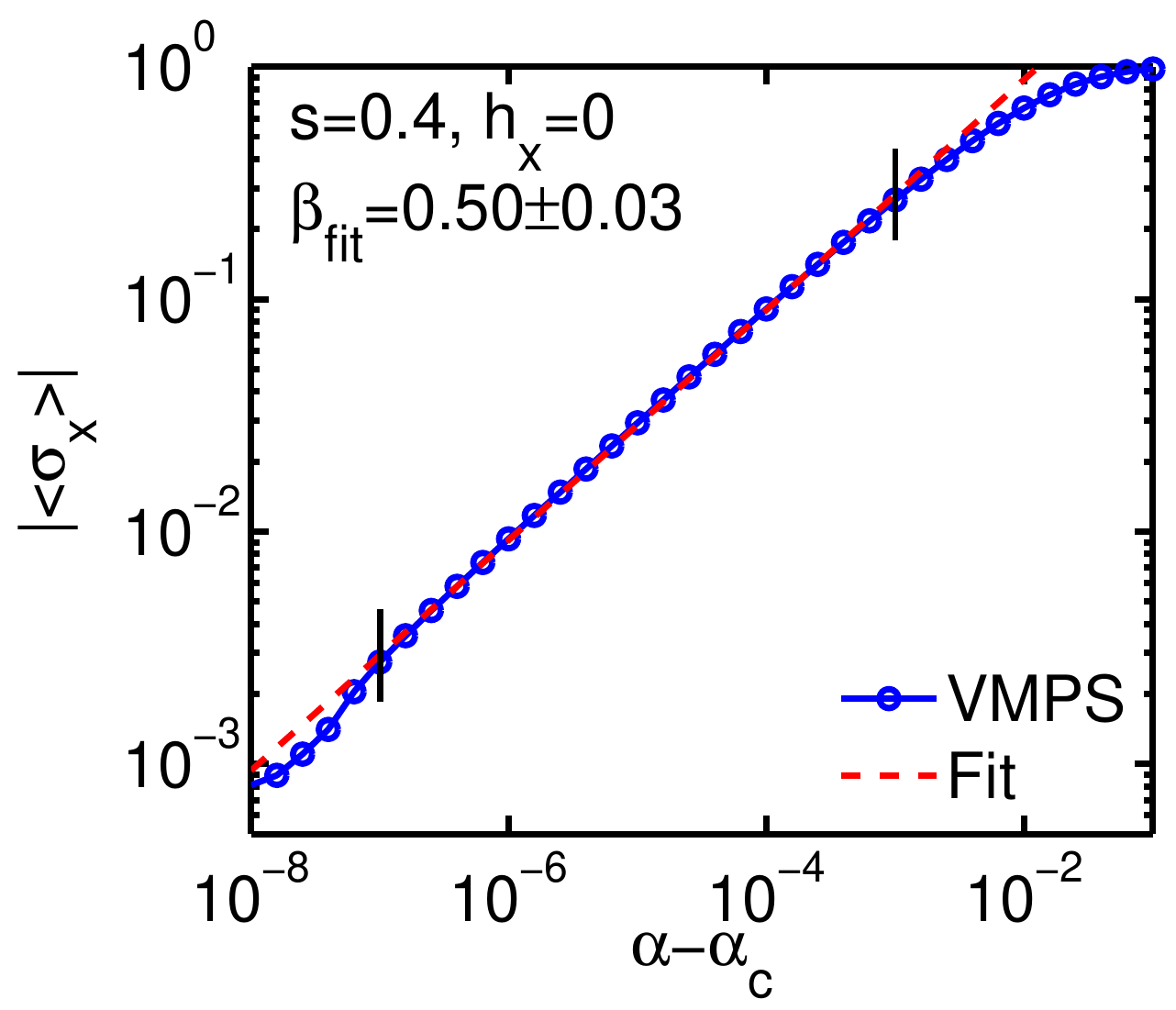}
\includegraphics[width=0.49\linewidth]{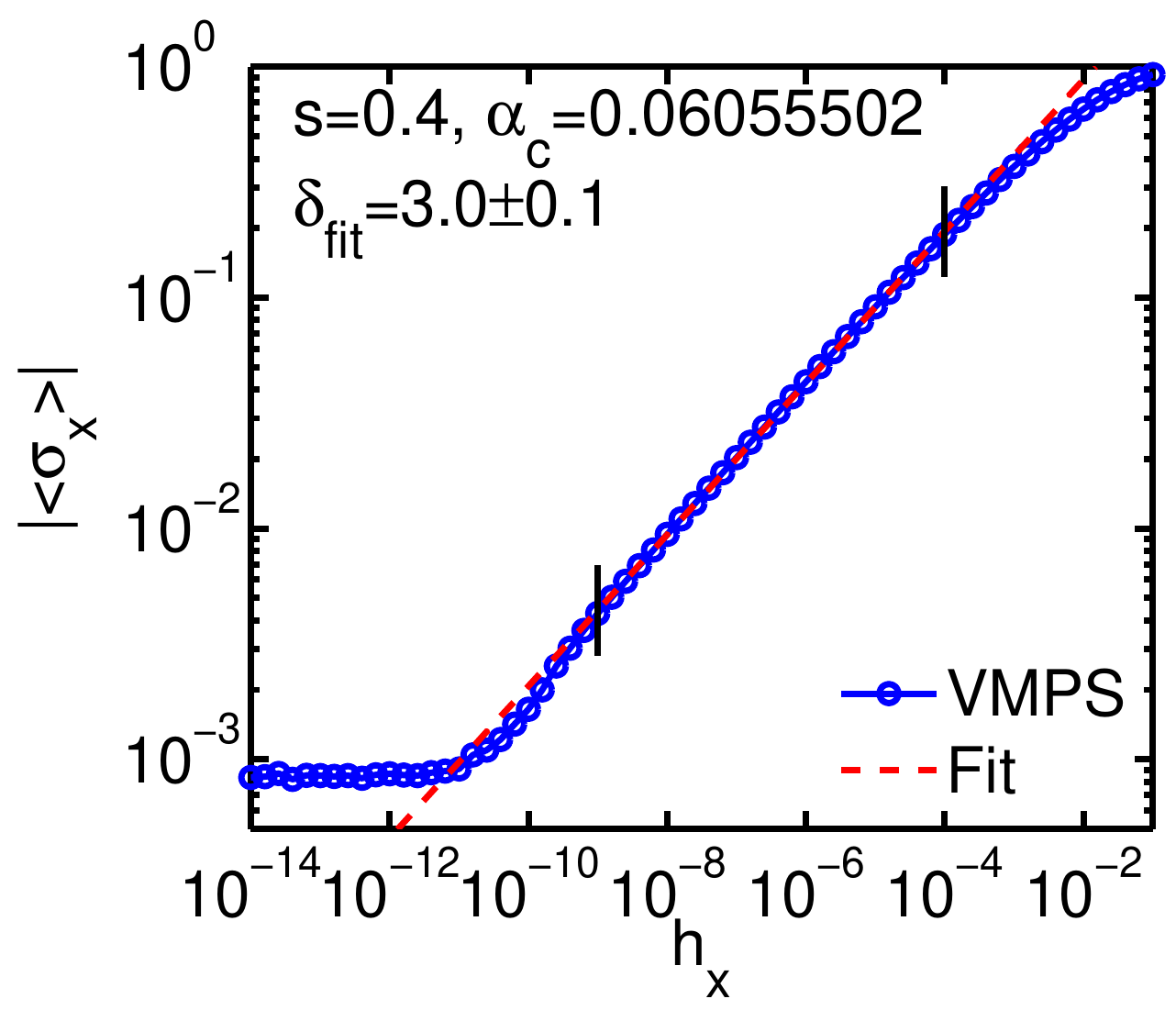}
\includegraphics[width=0.49\linewidth]{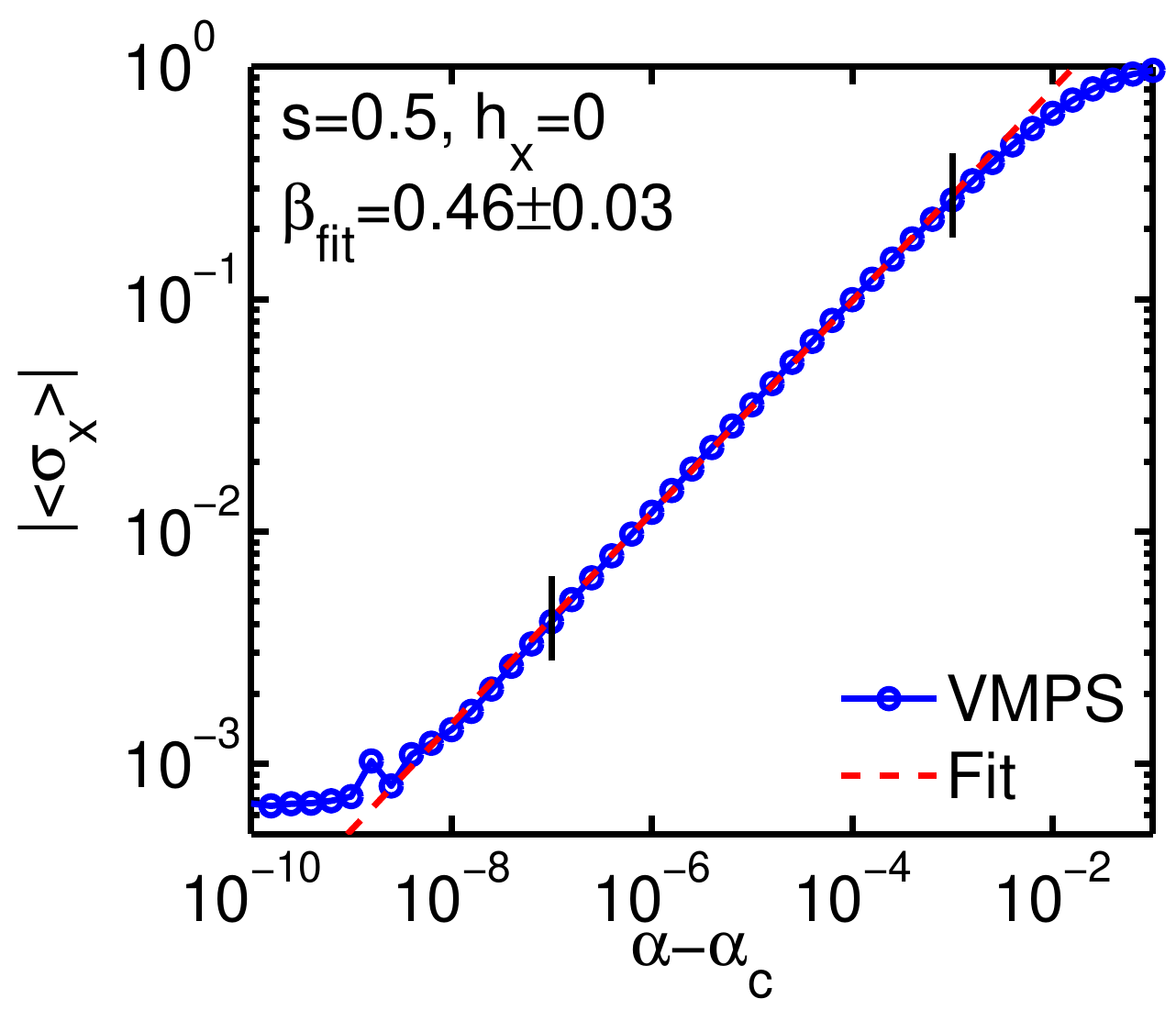}
\includegraphics[width=0.49\linewidth]{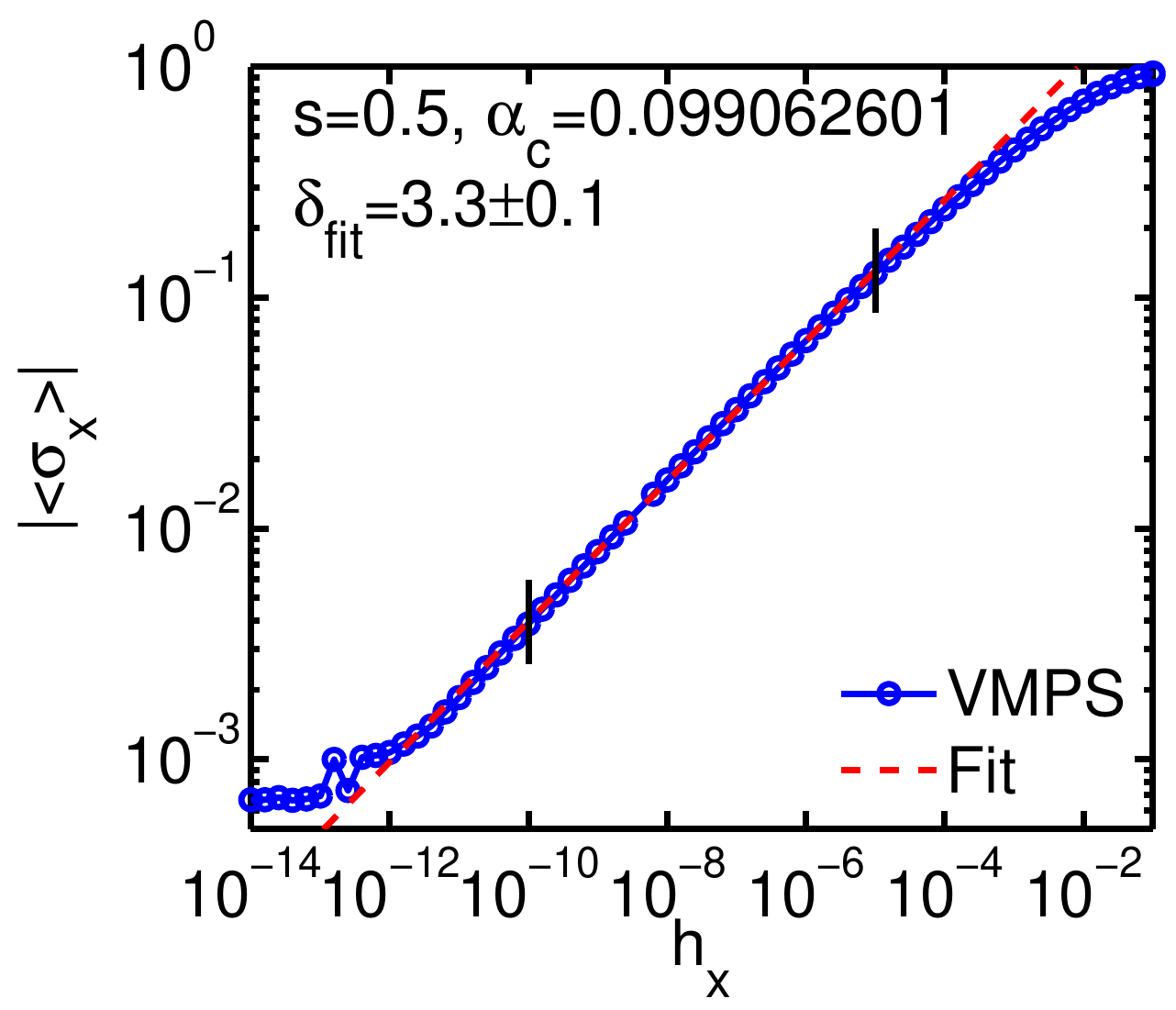}
\includegraphics[width=0.49\linewidth]{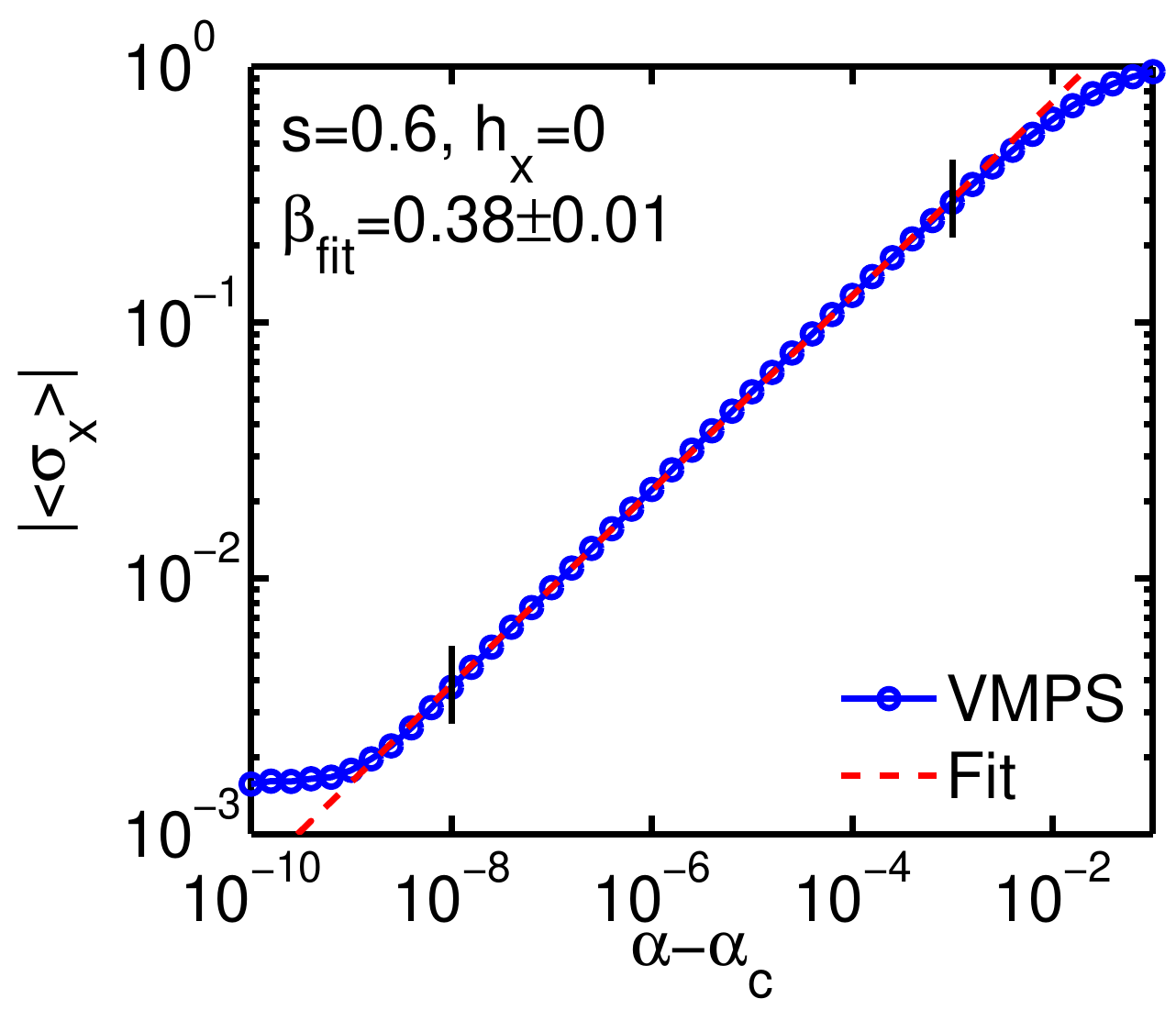}
\includegraphics[width=0.49\linewidth]{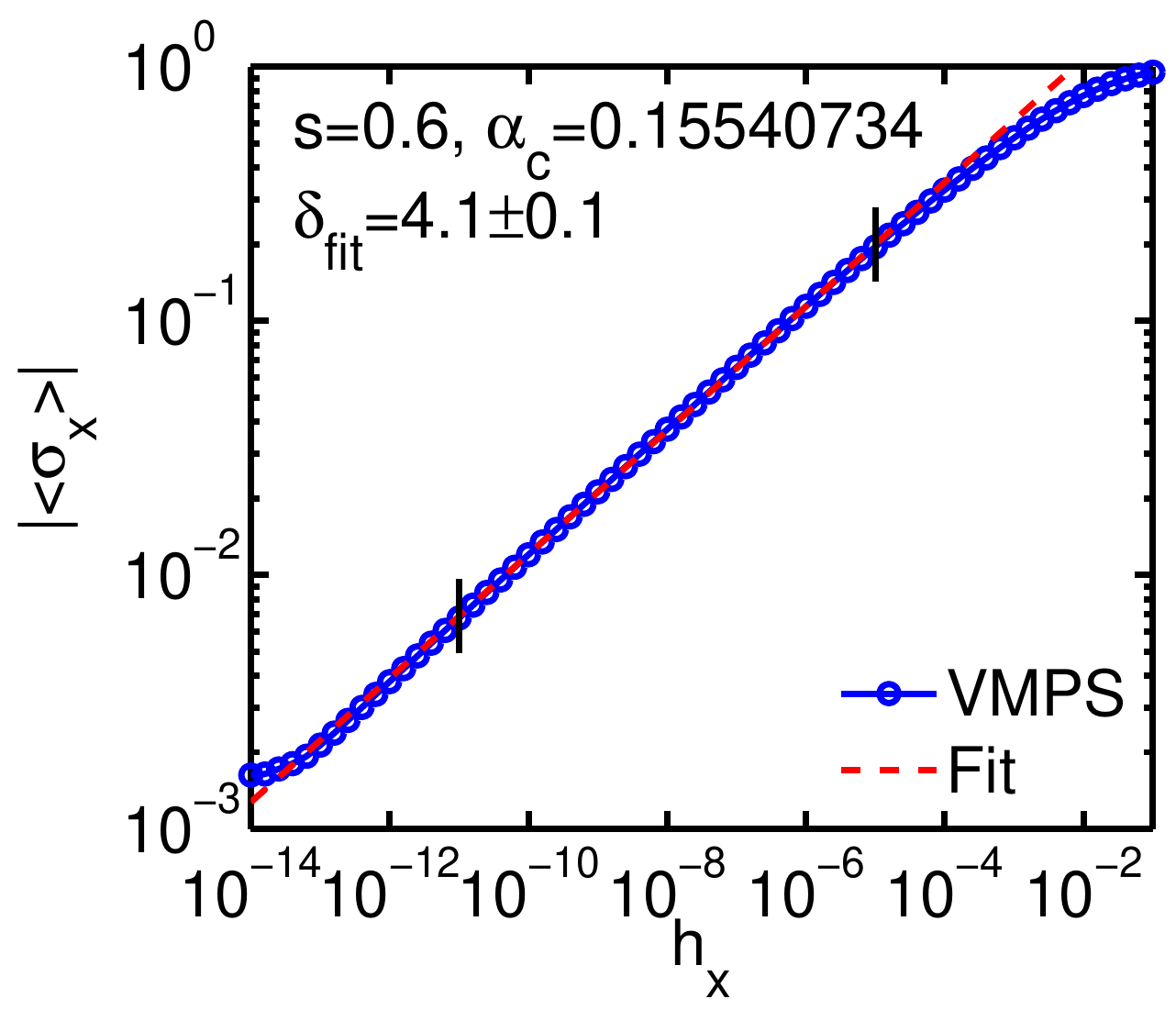}
\includegraphics[width=0.49\linewidth]{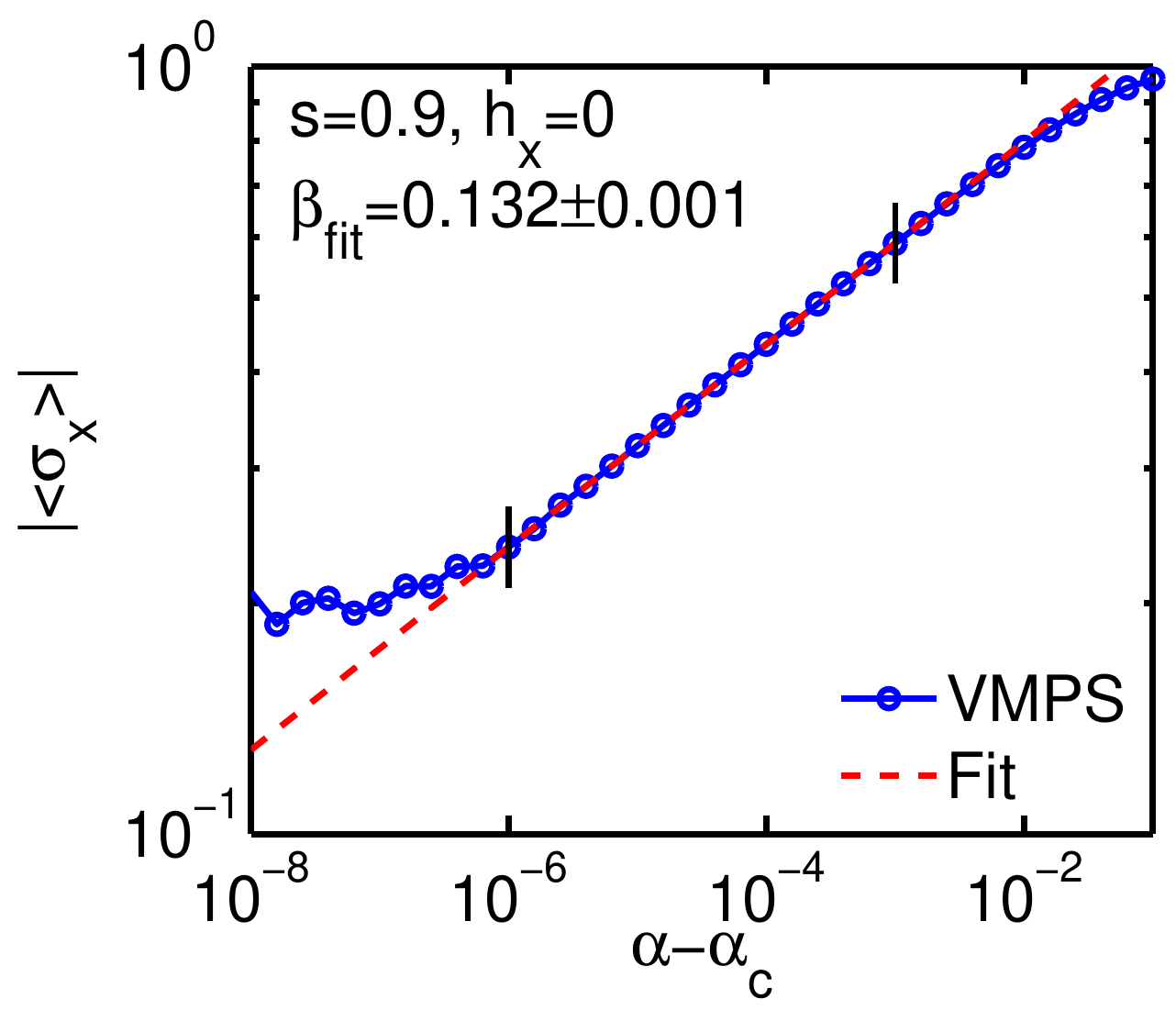}
\includegraphics[width=0.49\linewidth]{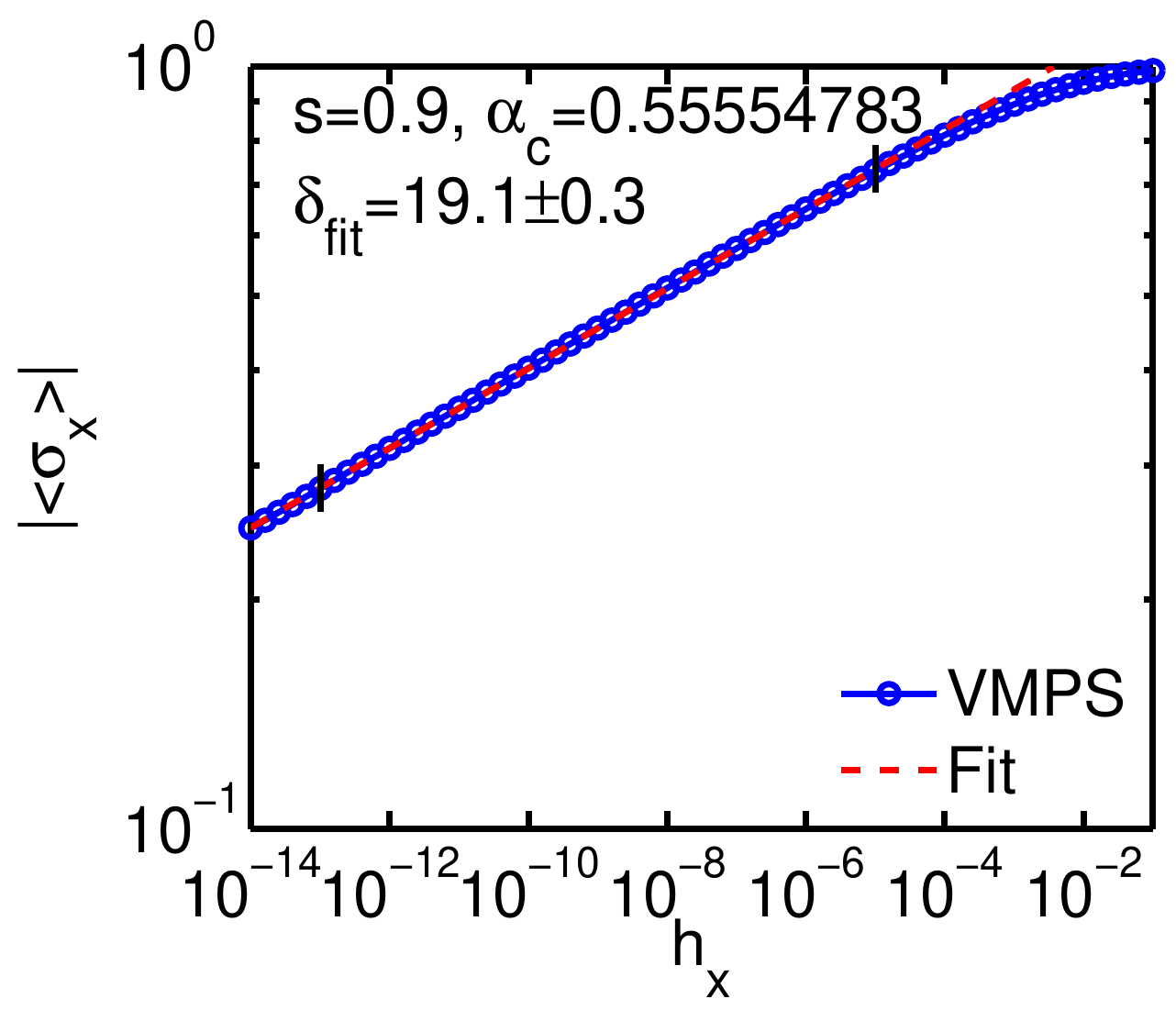}
\caption{The left and right columns show some of the VMPS results
used to determine the \onebm\ exponents $\beta$
and $\delta$ that are shown in Fig.~{\figfit}c) and
\figfit{d)} of the main text, respectively.
We used $h_{z}=0.1$, $\Lambda=2$, $L=60$, $D=40$,
  $d_\optimal=12$. The main text (\Eq{eq:requisite-chain-length})
explains why the range of pure power-law behavior is 
smaller for $s=0.2$ than for larger $s$-values. \vspace{-10mm}
\label{Flo:fig_fitting}}
\end{figure}

Fig.~\ref{Flo:fig_fitting} shows some typical data sets used to extract
the critical exponents $\beta$ and $\delta$ for $s<1$ shown in
Figs.~{\figfit}c,d  of the main text.  This allows to assess the
accuracy of the VMPS method in the quantum critical regime.

To properly describe the critical behavior, the Wilson chain must be
long enough to resolve energy scales down to the scale $T_\ast \propto
|\alpha - \alpha_c|^{\nu}$, which bounds the quantum critical regime
\cite{bulla_numerical_2003}; here $\nu$ is the correlation-length
exponent.
Now, the lowest energy scale accessible for a bosonic Wilson chain of
length $L$, i.e. $L -1 $ boson sites, is $\Lambda^{-L}$. Thus,
to determine $\alpha_c$ with an accuracy of, say, $10^{-a}$, we need
$\Lambda^{-(L-1)} \lesssim T_\ast \propto 10^{-a \nu}$, implying
that the requisite chain length scales as
\begin{eqnarray}
\label{eq:requisite-chain-length}
 L \sim a \nu \frac{\ln (10)}{\ln \Lambda} \; .
\end{eqnarray}
For \onebm, the correlation-length exponent becomes large for small
$s$, see Fig.~5a of Ref.~\onlinecite{bulla_numerical_2003}, and hence the
requisite chain length increases with decreasing $s$, too. Together
with Eq.~(\ref{eq:expshift}), according to which the largest shifts at
the end of the chain scale as $\Lambda^{L(1-s)/2}$, this implies that
in the localized phase, the sweep time needed to reach convergence
increases exponentially as $s$ decreases below 1/2.

For the above reasons, the data for $s=0.2$ is clearly less accurate
than for $s\ge0.3$, as reflected by the error bars shown in Fig.~2c,d of the
main text.  The two extremal values that define the indicated error
bars correspond to the two values of the exponent obtained by using
only the upper or lower half of the full fitting interval that is
indicated by vertical marks; the uncertainties from the straight
power-law fit are somewhat smaller.

For $s=1/2$, the transition is at its upper critical dimension and logarithmic
corrections to the leading power laws are expected. As quantifying these corrections from
the VMPS results is difficult, we have restricted ourselves to fits to effective power
laws, which then naturally result in exponents slightly deviating from the mean-field
values.

\subsection{Hyperscaling}
\label{sec:hyper}

For \onebm, a scaling ansatz for the singular part of the free energy
can be used \cite{VTB} to derive a hyperscaling relation between the
critical exponents $\delta$ and $x$, namely $\delta= (1+x)/(1-x)$.
Here, $x$ describes the divergence of the static susceptibility as
$T\to 0$ at criticality, $\chi(T) \propto T^{-x}$ where
$\chi=\partial\langle\sigma_x\rangle/\partial h_x$.
Hyperscaling also implies $x=y$, where $y$ characterizes the
divergence of the zero-temperature dynamic susceptibility,
$\chi(\w) \propto \w^{-y}$. Furthermore, $y=s$ is an exact
result for the critical long-range Ising chain at all $s$ (Refs.~\onlinecite{fisher,suzuki}).
This finally yields
\begin{equation}
\delta = \frac{1+s}{1-s}
\label{hyper}
\end{equation}
under the condition that hyperscaling is fulfilled, i.e., that no
dangerously irrelevant operators spoils the naive scaling
hypothesis. For the Ising chain, this applies below the upper critical
dimension, i.e., for $s>1/2$, while a dangerously irrelevant operator
appears for $s<1/2$. Consistent with the expectations from QCC, our
$\delta$-values follow the hyperscaling prediction \eqref{hyper} for
$s>1/2$ (see Fig.~2d of main text).

%%%%%%%%%%%%%%%%%%%%%%%%%%%%%%%%%%%%%%%%%%%%%%%%%%%%%%%%%%%%%%%%%%%%%

\section{Additional results for \twobm}

\subsection{Determination of phase boundaries}

\begin{figure}
\includegraphics[width=0.9\linewidth]{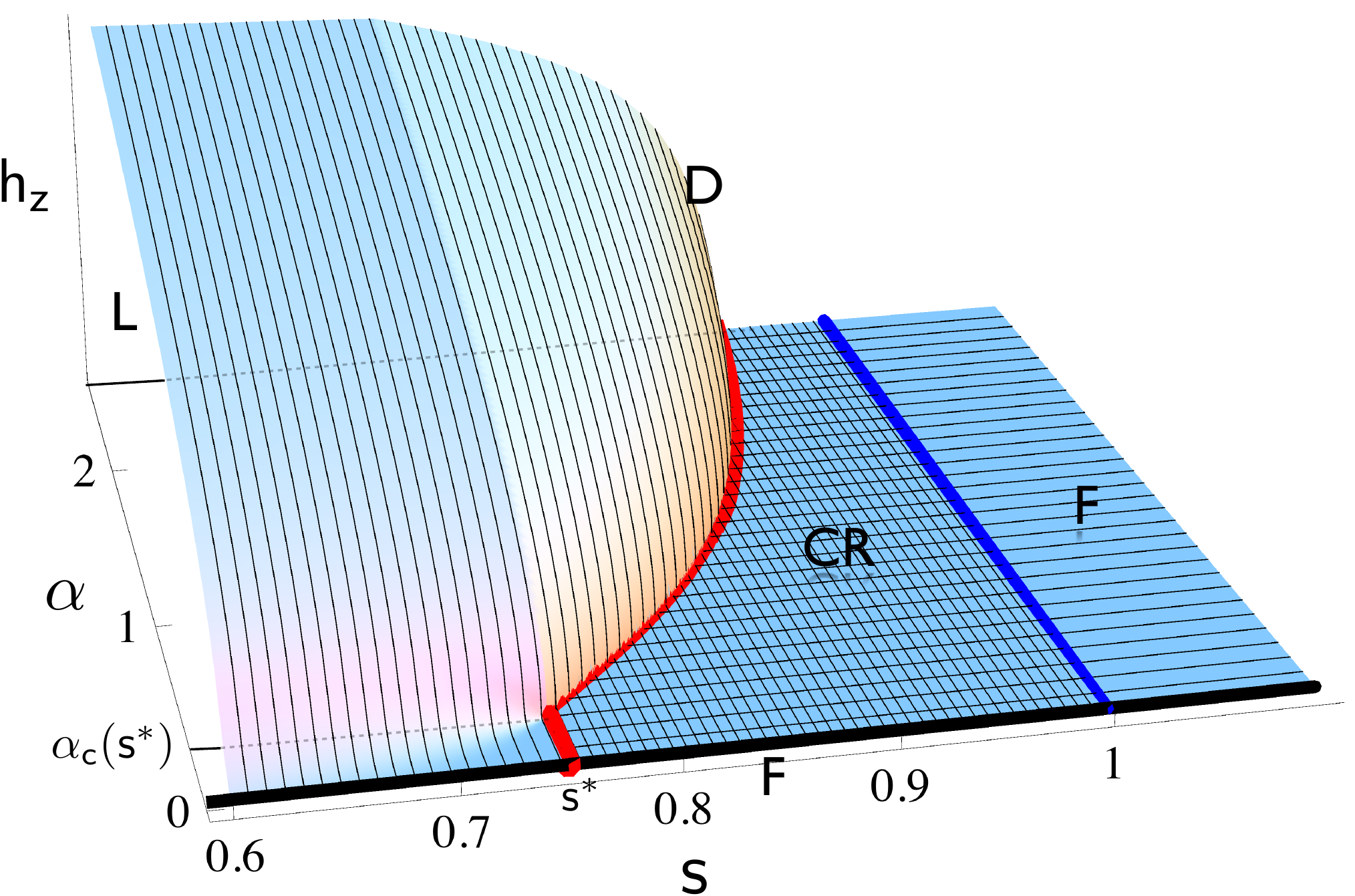}
\label{fig:3d-phase-diagram}
\caption{Schematic sketch of the \twobm\ phase diagram as function of
  $s$, $\alpha$ and $h_z$. The localized phase (L) lies below the
  curved surface representing the critical field $h_z^\critical
  (s,\alpha)$, the delocalized phase (D) above and to its right.  (The
  curved surface was not calculated, but is an artist's impression.)
  The cross-hatched area in the $h_z=0$ plane represents the critical
  phase (CR).  Its phase boundary with the localized phase (L) is
  marked by the thick red line in the $h_z=0$ plane (corresponding to
  the red line in main text, Fig.~3a), consisting of a straight line
  between the points $(s^\ast,0)$ and
  $(s^\ast,\alpha_\critical(s^\ast)$), and a curved portion
  representing the line $\alpha_\critical (s)$, for $s^\ast < s <
  1$. The thick blue line at $s=1$ is the CR-F phase boundary to the
  free phase (F).  The thick black line at $\alpha = h_z =0$ likewise
  represents a free spin (F).}
\label{fig:3D-boundary}
\end{figure}

Below we describe several approaches that we have
found useful for determining the various phase boundaries of
\twobm. For convencience, these boundaries are illustrated
schematically in \Fig{fig:3D-boundary}.

\begin{figure}
\includegraphics[width=0.95\linewidth]{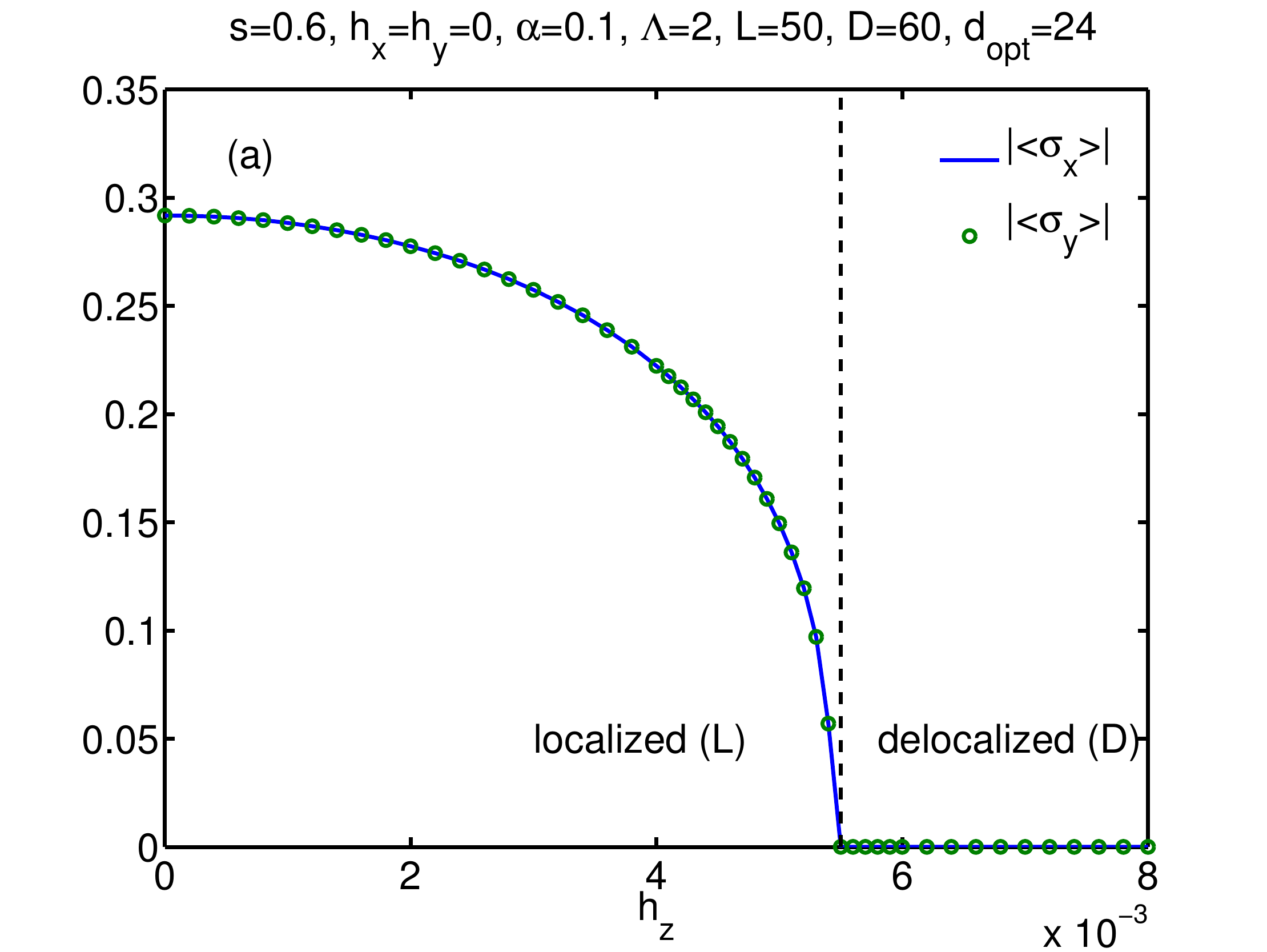}
\includegraphics[width=0.95\linewidth]{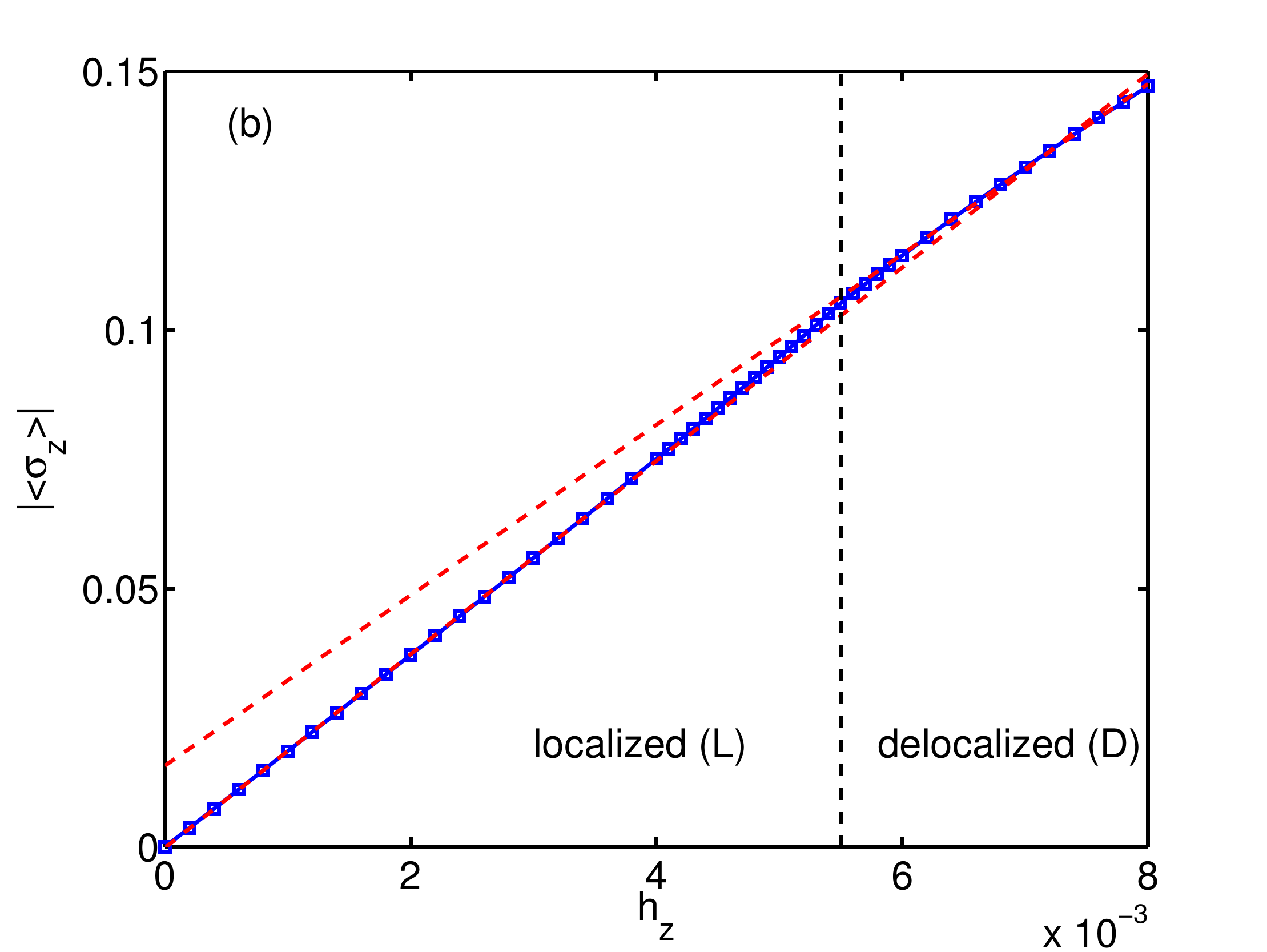}
\caption{
  (a) Order parameter $|\langle \sigma_{x,y} \rangle|$ for the L-D
  transition, driven to zero by increasing the transverse field $h_z$
  past the critical value $h^\critical_z$ (indicated by black dashed
  line).  (b) Correspondingly, the transverse-field response of
  $|\langle \sigma_z \rangle|$ shows a slight kink, indicating the
    higher-order singularity expected at the L-D transition. As in
    Fig.~3 of the main paper, the linear response of the L phase is
    clearly visible.
    We observe a small breaking of the XY symmetry of
    \twobm\ in the L phase due to numerical errors,
     in that the order parameter in panel (a) prefers configurations with
     $|\langle\sigma_{x}\rangle|
    = |\langle\sigma_{y}\rangle|$, instead of exploring the full
    rotational symmetry.
  }
\label{Flo:sxy_hz}
\end{figure}

  \emph{1. Order parameter.} Measurements of the order parameter
  $\langle\sigma_{x,y}\rangle$ of the localized phase can be used to
  determine the critical field $h_z^\alpha$ that defines
the location of the L-D boundary at finite $h_z$. This is
  demonstrated in Fig.~\ref{Flo:sxy_hz} for $s=0.6$ and
  $\alpha=0.1$. The method of ``best power laws'' is then suitable to
  obtain accurate values for $h_{z}^\critical$ as well as
  corresponding critical exponents -- we leave this for a future
  study.

Notably, the inherent instability of the CR phase, combined with
numerical errors of VMPS, renders difficult the accurate determination
of the CR-L phase boundary at $\vec{h}=0$.  In particular, a direct
observation of the order parameter as function of $\alpha$ or $s$
leads to sizeable uncertainties.

\begin{figure}
\includegraphics[width=1\linewidth]{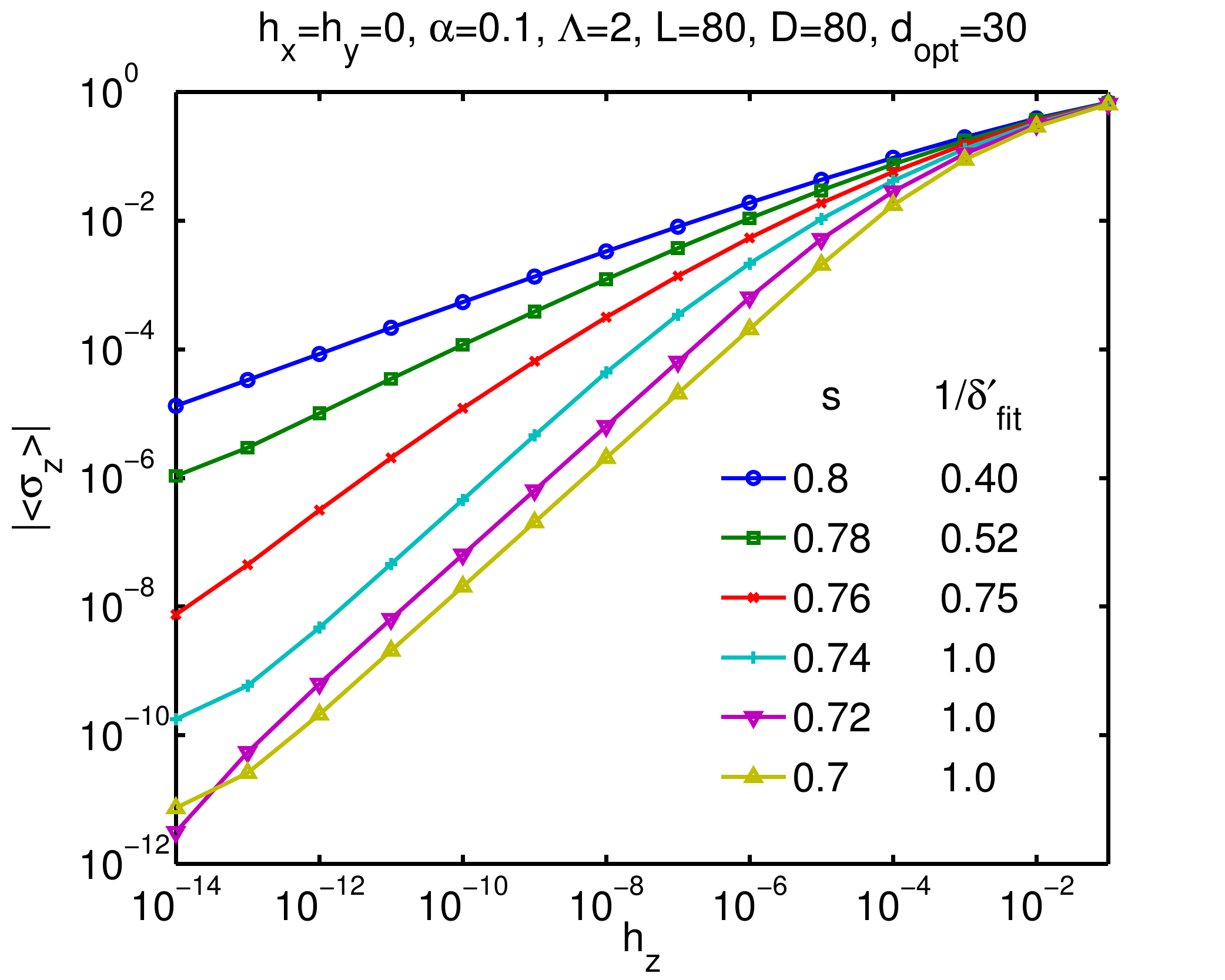}
\caption{Transverse-field response for \twobm, $|\langle \sigma_z
    \rangle |$ vs.\ $h_z$, used to determine the L-CR phase
    boundary. The fitting exponent $\delta'$ is defined in \Eq{pert}, and the fitting range is $h_{z} \in [10^{-12}, 10^{-8}]$.
    From the transition between a linear response for $s <
    s^\ast$ and a sublinear power law for $s > s^\ast$, we estimate
    $s^\ast$ to lie between $0.74$ and $0.76$.  To reduce numerical
    noise, the curves were calculated using a VMPS ground state that
    also was an odd eigenstate of the parity operator $\hat P_z$ of
    \Eq{eq:Parity-z-SBM2}.  }
  \label{Flo:transversefield}
\end{figure}

\emph{2. Transverse-field response.} The zero-field phase boundary
CR-L is most accurately extracted via the response to a small
transverse field, $\langle\sigma_{z}\rangle(h_z)$. The stable L phase
responds linearly, while the CR phase responds with a non-trivial
sublinear power-law, see Eq.~\eqref{deltap} below -- those can be
easily distinguished, as shown in Fig.~3b of the main paper. Indeed,
by studying the transverse-field response for small $\alpha$ and
$0.7\leq s\leq 0.8$, we are able to determine the value of the
universal critical ``dimension'' $s^\ast$ to be $0.75 \pm0.01$, as
shown in Fig.~\ref{Flo:transversefield}.

The accurate determination of $s^\ast$ requires the numerical
  results to be reliable down to transverse fields as small as $h_{z}
  \simeq 10^{-14}$.  Such precision is achievable, in principle, by
  using sufficiently large $D$ and $d_\optimal$, as follows from
  Fig.~\ref{Flo:fig_SBM2_conv}. However, such a brute force approach
  is computationally demanding. A more efficient strategy is to
  exploit parity symmetry: For $h_x = h_y = 0$ the \twobm\ Hamiltonian
  commutes with the parity operator
\begin{equation}
  \hat{P}_z \equiv \sigma_{z} e^{i\pi\hat{N}}
\text{,}
\label{eq:Parity-z-SBM2}
\end{equation}
where $\hat{N}=\sum_{k} \hat{b}_{k x}^{\dagger} \hat{b}_{k
  x}^{\phantom\dagger}+\sum_{k} \hat{b}_{k y}^{\dagger} \hat{b}_{k
  y}^{\phantom\dagger}$ counts the total number of bosons on the
entire Wilson chain. Thus, the ground state has a parity
degeneracy. In fact, this degeneracy is the main source of numerical
error for small $h_z$ when using a VMPS code that does not account for
parity symmetry, since then the degeneracy is
lifted by numerical noise (similarly to the \onebm\ case). By implementing this
parity symmetry explicitly in the code, we were able to achieve the required
precision (much better than in Fig.~\ref{Flo:fig_SBM2_conv}), while
using choices for $D$ and $d_\optimal$ that were not
unreasonably large, as shown in Fig.~\ref{Flo:transversefield}.

 \emph{3. Diverging boson number and flow diagram.}
We have also explored alternative ways
for determening the L-CR phase boundary,
based on monitoring the
the divergence of the boson number per site
at the end of the chain, or analyzing NRG-like
flow diagrams (as discussed for \onebm).
We have found these methods to be computationally
much cheaper than studying the transverse-field response,
while yielding results of comparable accuracy
for the L-CR phase boundary.

\begin{figure*}[t]
\includegraphics[width=0.48\linewidth]{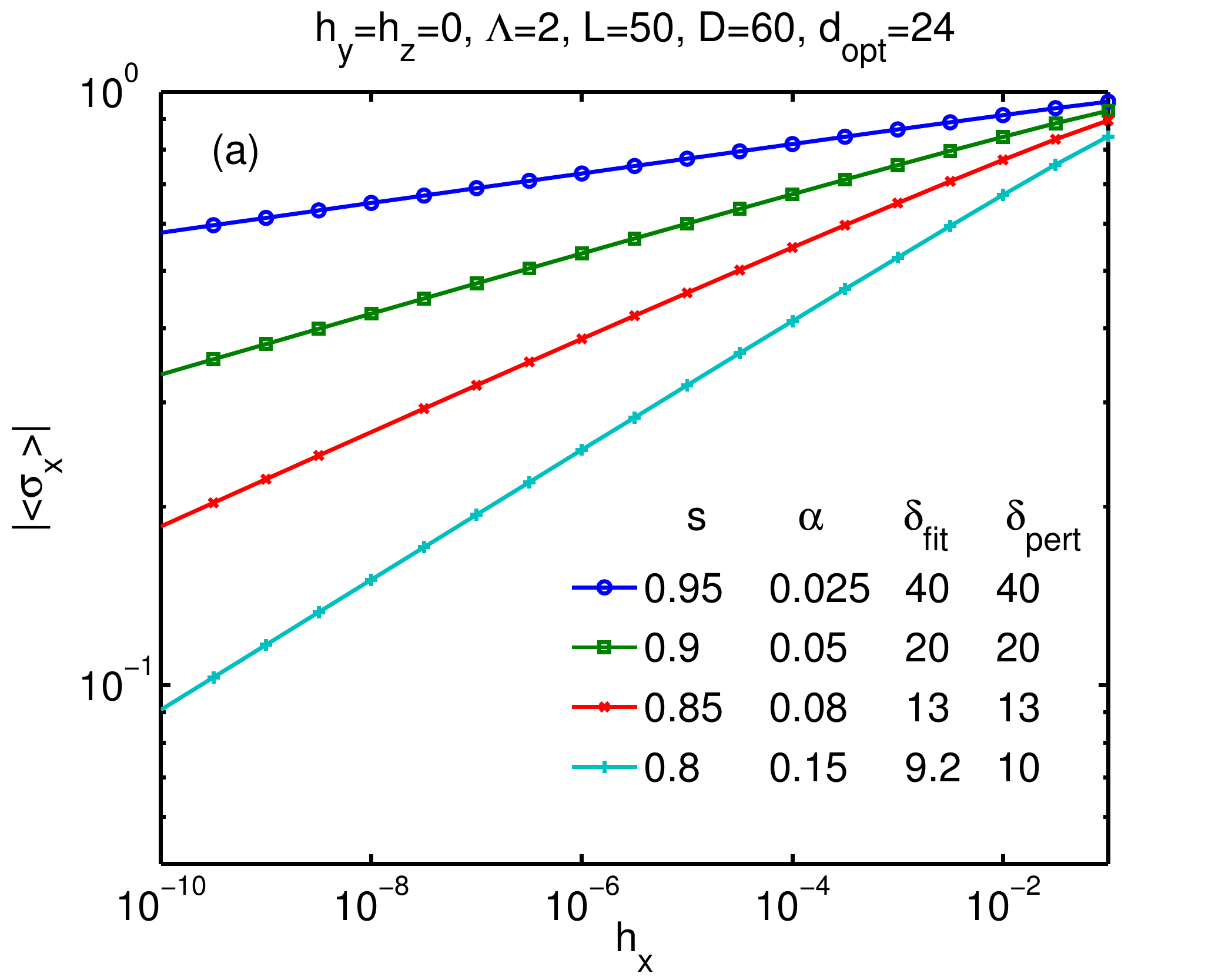}
\includegraphics[width=0.48\linewidth]{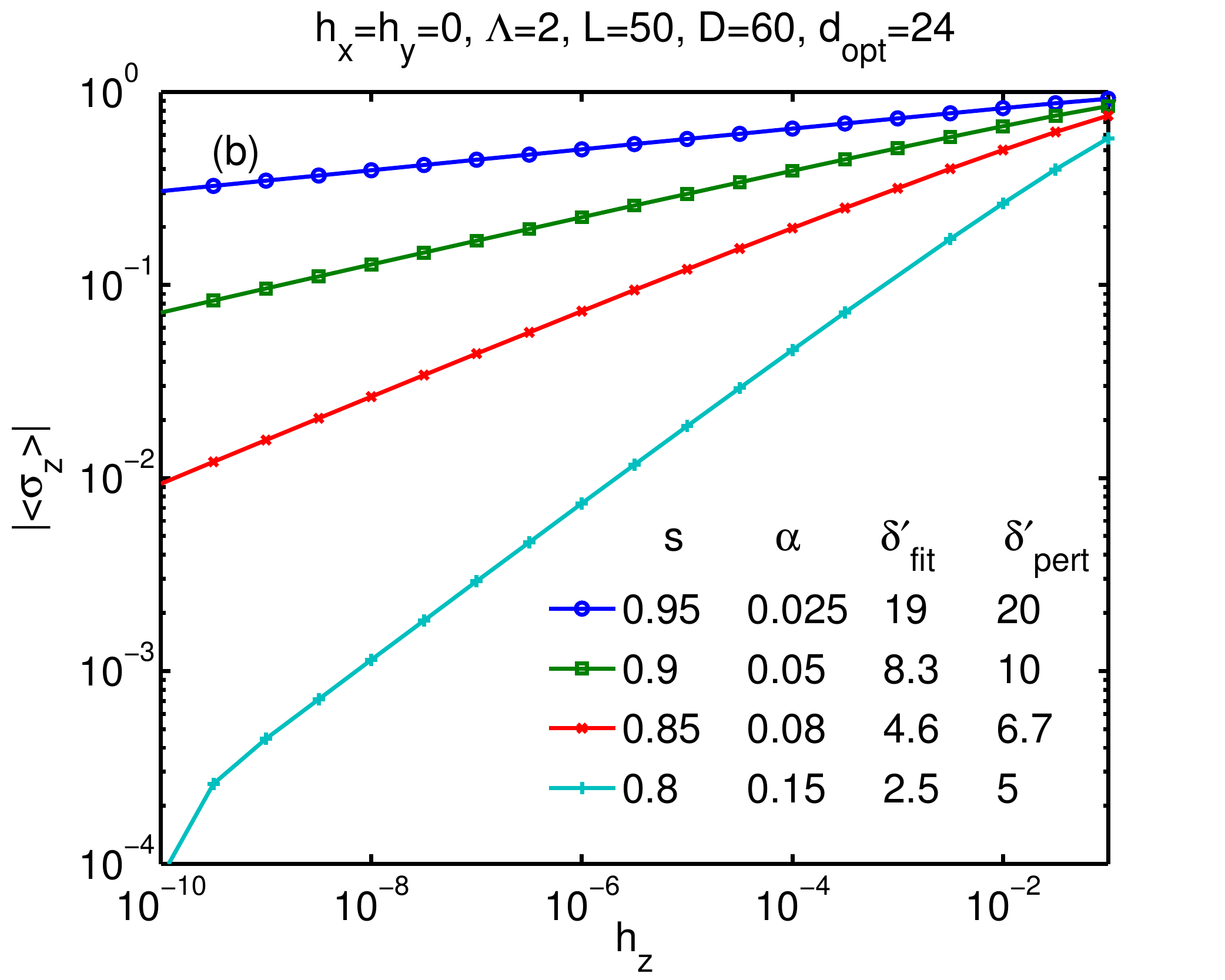}
\caption{Non-linear response \emph{inside} the CR phase. We find power
  law behavior as expected from \Eq{deltap}.  For each $s$, we
    chose an $\alpha$-value close to the $\alpha^{\ast}$ fixed-point value, to ensure
  that the asymptotic power law is reached quickly (i.e.\ without
    the need for $h_x$ to become extremely small).  Using the fitting
  range between $10^{-9}$ and $10^{-2}$, we find values for the
  exponents $\delta_{\rm fit}$ and $\delta'_{\rm fit}$ in
    excellent agreement with the values $\delta_{\rm pert}$ and
    $\delta'_{\rm pert}$ expected from \Eq{pert}, as indicated in the
  legends.  }
\label{Flo:SBM2delta}
\end{figure*}

\subsection{Properties of the critical phase}

  The CR phase corresponds to a partially screened (or fractional)
  spin, with non-trivial power-law autocorrelations of the components of
  $\vec{\sigma}$. In ground-state calculations, those can be probed by
  measuring the response to an applied field: The linear-response
  susceptibility at $T=0$ is infinite, and the non-linear response is
  of power-law character,
\begin{equation}
\langle \sigma_{x,y} \rangle \propto h_{x,y}^{1/\delta},~~
\langle \sigma_z \rangle \propto h_z^{1/\delta'} ,
\label{deltap}
\end{equation}
with $\delta,\delta'>1$. A standard renormalized perturbation expansion around the
free-spin fixed point results in \cite{rg_bfk}
\begin{eqnarray}
1/\delta  &=& \frac{1-s}{2} + \mathcal{O}([1-s]^2), \nonumber\\
1/\delta' &=& 1-s + \mathcal{O}([1-s]^2)\,.
\label{pert}
\end{eqnarray}

In Figs.~\ref{Flo:SBM2delta} we show numerical VMPS data for the non-linear response for
several sets of parameters inside the CR phase. We indeed find the expected power laws
over a sizeable range of fields; the power laws are cut off at small fields by the
influence of numerical errors, like the breaking of the XY symmetry of the model or the
generation of a small transverse field. From the data we can extract exponents $\delta$
and $\delta'$ as function of $s$. Comparing the resulting exponents to the perturbative
prediction \eqref{pert}, we find that the agreement is excellent, considering that
(i) for $s$ very close to unity the asymptotic regime is difficult to reach, due to
the slow flow of $\alpha$, and (ii) for larger $(1-s)$ the second-order terms in
Eq.~\eqref{pert} (i.e. two-loop corrections) will become important.

\subsection{RG flow near $s^\ast$}

Fig.~4 of the main text shows the RG flow of SBM2 in the two cases (a)
$s^\ast < s < 1$ and (b) $0<s<s^\ast$. In case (a) the CR phase is
stable for small $\alpha$ and $\vec{h}=0$, whereas this phase is
absent in case (b). In the following we briefly discuss the evolution
of the RG flow from case (a) to case (b), as this is related to a 
\textit{discontinuous} jump of the phase boundary upon varying $s$.

As described in the main text, the fixed points CR and QC1 of case (a)
approach each other upon lowering $s$, such that they meet at
$s=s^\ast$ and both disappear for $s<s^\ast$. This follows from the
absence of the CR phase in case (b) and the fact that both are
intermediate-coupling fixed points for $s<1$ (i.e. they have to meet
at finite $\alpha$). The merger of the two fixed points also implies
that the flow lines to the left and right of CR in Fig.~4a merge. As a
result, the RG flow line which lead from F via CR to D in case (a) now
becomes a flow line from F to QC2 in case (b). For the phase boundary
of the L phase (thick line in Fig.~4) this means that its starting
point on the $h_z=0$ axis jumps from a finite
value (corresponding to QC1) in (a) to zero in (b)
once $s$ is lowered past $s^\ast$.
%upon variation of $s$.

We note that the merger and disappearance of two fixed points upon
variation of a ``dimension'' $d$ is not unusual. For instance, upon
approaching the lower critical dimension $d_c^-$ of a magnet from
above, the critical fixed point typically approaches the trivial fixed
point describing the ordered phase, such that both merge at $d=d_c^-$
and disappear for $d<d_c^-$. Here the evolution of the phase boundary
is continuous.
What is unusual about SBM2 is that two {\em intermediate-coupling}
fixed points merge and disappear, causing the discontinuous
behavior. The only other example with similar physics we are aware of
is in the two-channel Anderson/Kondo impurity model with power-law
density of states $\propto |\omega|^r$, where a critical fixed point
merges with a stable non-Fermi liquid fixed point at some critical
dimension $r_{\rm max}$, with a consequent jump in the phase
diagram.\cite{pg2ck,Schneider11}

%%%%%%%%%%%%%%%%%%%%%%%%%%%%%%%%%%%%%%%%%%%%%%%%%%%%%%%%%%%%%%%%%%%%%